# Gerberto e la misura delle canne d'organo

**Costantino Sigismondi**

Sapienza Università di Roma e ICRA
International Center for Relativistic Astrophysics
Piazzale A. Moro, 5 00185 Roma Italia
costantino.sigismondi@gmail.com

**Abstract**
Gerbert of Aurillac in the *Mensura Fistularum* explained how to compute the length of organ pipes. The method is shown on two octaves, starting from a *fistula* of length L=16 units and radius 1 which is equivalent at a monochord of length λ=18. The adopted acoustic correction for the first octave to the Pythagorean lengths is L=λ-α·r with α=2.
The lower octave starts from L=36-2=34 units. The proportion 16:34=34:x is used for obtaining the next diapason. All lengths of the notes of this second octave follow this proportion and no more the additional acoustic correction.
Gerbert finds the same multiplicative law for computing pipes and monochord's lengths, opportune constants allow to switch from monochord (12) to lower organ octave (14+1/3+1/144+1/288) to the higher one (13 + ½ ).
The purpose of this treatise is to show the same mathematical order, given by God, below different acoustical phenomena. This is a modern perspective in history of science, because experimental data (practical acoustical corrections) are also taken into account. The treatment is limited to pipes of same diameter, and it is no conceived for organ builders. An Italian translation of the core text of the *Mensura Fistularum* is offered.
The experimental measurement of *end* and *mouth* corrections for two pipes of different forms and for the flute is presented to support with modern acoustics approach that discussion.

**Sommario**
Nella *Mensura Fistularum* Gerberto ha spiegato come calcolare la lunghezza delle canne d'organo. Il metodo è esemplificato su due ottave: per la prima ottava la nota fondamentale corrisponde alla più acuta della serie e viene scelta una *fistula* lunga 16 unità e di raggio 1, ed equivale ad un monocordo di riferimento lungo λ=18 unità. In un solo manoscritto c'è la corrispondenza tra questa lunghezza iniziale e quella di un'ulna. La correzione da applicare alle *fistulae* di lunghezza L e raggio r della prima ottava è λ-L=α·r, con α=2, dove λ è la lunghezza equivalente al monocordo che segue esattamente le proporzioni pitagoriche tra le lunghezze come 9/8 per scendere di un tono, 4/3 per un intervallo di quarta, 3/2 per una quinta e 2 per un'ottava, che risulta pari a L=34 unità, ma suona come fosse un monocordo lungo λ=36.
Questo valore di α, noto solo per via empirica al tempo di Gerberto, include le correzioni di bocca e di apertura libera della *fistula*, ed è plausibile con i dati sperimentali, ma non è costante con la frequenza del suono, e si annulla per frequenze alte, corrispondenti a *fistulae* piccole. Qui espongo il metodo per misurare sperimentalmente α in funzione della frequenza analizzando lo spettro dei suoni armonici prodotti da tubi.
Per la seconda ottava, quella inferiore, Gerberto si è basato sulla proporzione geometrica 16:34=34:*x*, dove *x*=72 ¼ è la lunghezza della *fistula* maggiore, che dovrebbe suonare due ottave sotto la nota di partenza, ed invece suona un quarto di tono più bassa come λ=74 ¼ perché dal punto di vista della fisica acustica moderna continua a valere la correzione con α=2.
Se il numero 13 ½ è il fattore costante per cui moltiplicare le differenze di lunghezze tra *fistulae* della prima ottava per frazioni pitagoriche, ad esempio per fare il salto di un'ottava occorre allungare la *fistula* di 18 unità cioè 34-16=18= 4/3 ·13 ½, e 4/3 è la frazione pitagorica.  Per la seconda ottava il numero diventa 14 +1/3+1/144+1/288 e la frazione pitagorica raddoppia.
Con questo algoritmo Gerberto mostra come intendere la commensurabilità tra *fistulae* di uguale diametro e monocordo, che altrimenti *non conveniant*, compilando una tabella che copre due ottave e parte dal numero 2304 (comune ai tre generi della musica antica) fino al 10404 dove le proporzioni sono le stesse ricavate nell'intervallo tra 16 e 72 ¼ . Il fine è chiaramente didattico, essendo Gerberto ben al corrente del fatto che le canne in un organo hanno diametri differenti.
È il primo tentativo, in acustica, di *salvare i fenomeni* e le proporzioni pitagoriche mediante parametri correttivi mantenendosi in continuità con il mondo classico, l'equivalente degli equanti tolemaici in astronomia. La traduzione italiana dei passi centrali della *Mensura Fistularum* è qui offerta.

**Résumé**
Gerbert d'Aurillac dans la *Mensura Fistularum*, a donné une méthode pour calculer la longueur des tuyaux d'orgue à partir des proportions Pythagoriques. Il considère tuyaux de même diamètre et ça montre qu'il s'agit d'un essay didascalique, pour montrer qu'il y a un seul ordre mathématique dans la nature, donné par Dieu, et pas d'instructions pour la construction d'orgues.
Il part d'un tuyau long L=16 unités de et diamètre 2, et il sonne comme un monochorde de longueur $\lambda$=L+2·$\alpha$, avec $\alpha$=2. Ensuite il appliques cette lois à toutes les longueurs Pythagoriques de la première octave. Ainsi l'octave plus basse commence de L=34 et d'après la proportion géométrique 16 :34=34 :*x* il trouve la longueur de la note plus basse. La même proportion est utilisée pour calculer les longueurs de la seconde octave. La correction additive L=$\lambda$-2·$\alpha$ n'est plus utilisée. Gerbert avait trouvé une lois multiplicative qui permettait de calculer soit les longueurs d'un monochorde (avec le numéro 12) soit l'octave baisse (14+1/3+1/144+1/288) que l'octave acute (13+ ½ ). La traduction en Italien des parties centrales de la *Mensura Fistularum* est ici présentée.
Une discussion et des expérimentes d'acoustique moderne avec des tuyaux différents et un flute permit de compléter cette introduction sur les tuyaux d'orgue aussi du point de vue de la physique. Un tuyau simple ouvert des deux côtés admit une correction $\alpha$=1.2. S'il y a aussi une bouche où le son se produise cette correction peut arriver au valeur utilisé par Gerbert.
L'intention didactique de Gerbert est aussi claire de sa choix d'utiliser la proportion géométrique pour la seconde octave et pas la correction additive, qui aurait été plus précise acoustiquement.
Celui de Gerbert est le premier tentative en acoustique de adapter la description Pythagorique aux phénomènes, en utilisant données réels,  une perspective moderne dans l'histoire de la science.

## *Introduzione*

Gerberto *musicus* è stato studiato dal punto di vista sia della storia della musica[1] che per i suoi *scholia* sulla *Musica* di Boezio[2] e come costruttore di organi.[3]
Gli studi di Klaus-Jurgen Sachs hanno permesso di identificare l'origine gerbertiana dei trattati sulla *Mensura Fisutlarum*  grazie all'attribuzione a Gerberto del trattato *De Commensuralitate fistularum et monocordi, cur non conveniant* esplicitata nel manoscritto del XII secolo Ms 9088 f. 125-128 alla Biblioteca Nacional di Madrid,  testo prima di allora attribuito a Bernellino[4] di Parigi, autore che è posteriore a Gerberto di quasi un secolo. Sachs ha anche affrontato il problema delle correzioni acustiche alla lunghezza delle canne d'organo. [5] A Gerberto costruttore di organi ha rivolto la sua attenzione anche Clyde Brockett[6] che ha evidenziato il legame tra il *Carme figurato* di Gerberto del manoscritto Ms Latino 776 della Bibliothèque Nazionale di Parigi che accompagnava il dono di un organo all'imperatore Ottone II, includendo anche i le cifre indo-arabe nel secondo livello di criptazione, e Flavio G. Nuvolone[7] ha portato a piena maturazione questa ipotesi nel corso degli ultimi 10 anni. Interessante anche la prospettiva ripresa dalla Harriett Pratt Lattin che nel commentare un passo della lettera al monaco Bernardo di Aurillac (n. 105) annota a proposito del passo *Vel in his quae fiunt ex organis* che questo suggerisce la nuova tecnica –con l'uso delle dita- per suonare l'organo che si stava sviluppando a quel tempo, che forse vedeva la luce proprio con Gerberto. Prima si usava tutta la mano per tirare i tasti.[8]

---

[1] K. J. Sachs, *Mensura Fistularum,* Stuttgart (1970-1980); C. Meyer, *Gerbertus Musicus*, Aurillac (1997); E. Santi, *Gerberto e la Musica*,  Roma (2003).
[2] Lettere 4 e 5 a Costantino di Fleury; secondo la numerazione dell'edizione di  H. P. Lattin, *The Letters of Gerbert*, New York (1961).
[3] La lettera 77 all'Abate Geraldo del 20 febbraio 986 e la 102 all'Abate Raimondo di Aurillac del 24-31 gennaio 987 riguardano la promessa di un organo, e poi l'incertezza sulla sua sorte per le vicende politiche italiane.
[4] Cfr. PL 151, col. 0653 A e ss. BERNELINI CITA ET VERA DIVISIO MONOCHORDI IN DIATONICO GENERE, datato nel secolo XI. Nella nota introduttiva
[5] Sachs, *op. cit.,* vol. II p. 66 (1980) dove è riportata la formula l'=(0.33+kd/$\delta$)d per la lunghezza efficace di una *fistula*.

In questo testo si vuole proporre la traduzione italiana di un passo del manoscritto madrileno edito dal Sachs, quello riguardante la costruzione della prima e della seconda ottava, che inizia con le parole *Data Igitur*. Si pone attenzione all'algoritmo seguito da Gerberto in queste operazioni, cioè una vera e propria correzione per la prima ottava ed una proporzione armonica per ottenere la seconda ottava. Si aggiungono degli esperimenti di fisica acustica, per una misura empirica delle correzioni di bocca e di apertura libera di una *fistula*, per completare il panorama di una introduzione completa alla fisica delle canne acustiche.

## *Cenni sulla Teoria musicale nei secoli*

Nelle arti liberali del Quadrivio la musica sta all'aritmetica come l'astronomia sta alla geometria. Infatti sia la musica che l'astronomia sono l'implementazione in natura delle proporzioni matematiche e geometriche che si possono considerare mediante i numeri.

Una prospettiva, questa, di origine pitagorica, come recita il frammento di Filolao (attivo a Taranto nel V sec. a. C.) "Il Numero è l'autogenito vincolo dell'eterna stabilità delle cose cosmiche[9]".

La tradizione pitagorica fu ripresa nel periodo ellenistico: su incarico di Aristotele (384-322 a. C.) Aristosseno scrisse una storia della musica andata perduta.

Cicerone (100-43 a.C.) trattò alcuni argomenti nel Somnium Scipionis, ripresi da Posidonio (135-51 a.C.) che fu suo maestro.

Il grande astronomo Tolomeo (90-168 d. C.) scrisse l'*Armonica*, in tre libri, e secondo il commento di Porfirio si basò su uno scritto di Didimo sul contrasto tra la teoria pitagorica e quella di Aristosseno.

Macrobio commentò il Somnium nel IV sec. d. C., e Calcidio, enciclopedista del IV sec. d. C., nel suo commento al Timeo ne tramandò i temi al Medioevo.

Boezio (480-525 d. C.) lasciò il *De Institutione Musica* dove distinse la Musica in *Mundana, Humana e Instrumentalis*, come forme via via meno nobili. Pur tuttavia Gerberto si occupa anche della Musica Instrumentalis, mostrandone la derivazione da leggi pitagoriche anche per le canne dell'organo, che sembravano sfuggire ad una codifica per così dire *canonica*. Gerberto ebbe modo di spiegare alcuni passi del De Istitutione Musica ai suoi studenti, come ci testimoniano le lettere a Costantino di Fleury che risalgono al 978-980[10] e quanto riporta Richero di Reims nella sua biografia di Gerberto[11].

La ricerca di consonanze cosmiche, l'ideale boeziano di *Musica Mundana*, ha poi animato perfino la ricerca di Keplero (1571-1630), Matematico Imperiale di Rodolfo II († 1612), che a partire dalla sua prima opera nel 1595, il *Mysterium Cosmographicum* fino all'*Harmonices Mundi* (1619) ravvisò finalmente quei rapporti numerici che descrivono le armonie prodotte dalle orbite planetarie come se i loro semiassi fossero delle corde vibranti. Le sfere cristalline avevano ormai perso la loro realtà fisica diventando eteree, poiché Tycho Brahe aveva dimostrato che le comete le attraversavano senza subirne conseguenze, tuttavia Keplero non smise di cercare i numeri, preferibilmente interi e armonici, che le governavano.

---

[6] C. Brockett, *The Frontispiece of Paris, BN Ms Lat. 776. Gerbert's Acrostic Pattern Poems,* Manuscripta **39**, 3-25 (1995).

[7] F. G. Nuvolone, *Gerberto e la Musica*, ABob St **5** Bobbio 2005; *Appunti e Novità sul Carmen Figurato di Gerberto d'Aurillac e la sua Attività a Bobbio*, ABob **25** Bobbio 2003; *La Presenza delle Cifre Indo-arabe nel Carmen Figurato di Gerberto,* ABob **24** Bobbio 2004.

[8] H. Pratt Lattin, *op. Cit.*, p. 141 not. 5 che richiama a sua volta il testo di W. Apel, *The Early History of the Organ*, Speculum **23** 215-6 (1948).

[9] M. Timpanaro Cardini, *Pitagorici, Testimonianze e Frammenti*, 3 voll. La Nuova Italia, Firenze (1958).

[10] La quarta e la quinta lettera nella numerazione di Harriet Pratt Lattin [1961].

[11] Cfr. Laura Paladino, *La biografia di Gerberto nella Historia Francorum di Richero di Reims*, ABob **27-28** (2006).

L'organo idraulico fu inventato ad Alessandria d'Egitto da Ctesibio nel II sec. a. C..
Il peso dell'acqua serviva a mantenere in pressione l'aria che poi fluiva attraverso le canne.
È stato trovato nel maggio 1931 dal prof. Lajos Nagy l'unico esemplare di *hydraulis* greco-romano in uno scavo in Ungheria[12] nel sito di *Aquincum*, l'antica Buda, capitale della provincia romana della Pannonia Inferiore.
Pipino il Breve, nel 757, ricevette in dono dall'imperatore Costantino Copronimo un organo (*venit organum in Franciam* dicono le cronache) identico a quello che Bisanzio utilizzava già da qualche secolo e che quest'organo era pneumatico. Circa 50 anni più tardi, Carlo Magno ebbe in dono uno strumento analogo dal Califfo di Bagdad. Di altro organo si ha notizia verso l'anno 826, quando, da Ludovico il Pio, re dei Franchi, fu accolto il prete veneziano Giorgio che era un ammirabile costruttore *organorum cincinnandorum* e che ne costruì uno meraviglioso ad Aquisgrana.
Un altro ricordo si ha tra gli anni 872 e 882 quando il Papa Giovanni VIII scriveva ad Annone Vescovo di Frisinga: *precamur autem ut optimum organum cum artefice qui hoc moderari et facere ad omnem modulationis efficaciam possit ad instructionem musicae disciplinae, nobis aut deferas aut ... mittas*. Ma il più grande organo di cui sia rimasta memoria prima del Mille fu quello di Winchester (957) per suonare il quale erano necessari due organisti che, quantunque lo strumento contasse solo quattrocento canne e quaranta tasti, dovevano adoperare i pugni e i gomiti (donde il *pulsantur organa* del linguaggio liturgico); per azionare i ventisei mantici occorrevano settanta uomini. Da ciò si può arguire che gli organisti di questi tempi (*organum pulsatores*) dovevano essere scelti non per il loro talento musicale, ma per la loro forza fisica.
A partire dal secolo X l'uso dell'organo si generalizza nell'Occidente cristiano: non però organi monimentali, ma piccoli, detti *portativi* o *ninfali*: constano di una cassetta rettangolare con le canne disposte su di una, due o tre file normalmente decrescenti da sinistra verso destra con ai loro piedi la tastiera da nove a dodici tasti. Qui il suonatore è anche il tiramantici. Successivamente si diffondono gli organi *positivi* che si mettevano sopra un mobile o per terra, avevano una tastiera più estesa da suonarsi con due mani, mentre i mantici venivano azionati da altra persona.[13]

## *Scala Pitagorica e Fisica del Monocordo*

La scala musicale Pitagorica è basata sul funzionamento del monocordo.
Una corda bloccata ai suoi estremi, vibrando nel suo modo fondamentale suona una determinata nota. Se la corda viene bloccata ad una lunghezza diversa la nota sarà differente, ma se l'armonica fondamentale è lunga L e corrisponde alla nota do della prima ottava, si può ottenere un sol sempre della prima ottava salendo di un intervallo di quinta facendo vibrare 2/3 della lunghezza della corda.
Vale, per la scala diatonica, la seguente tavola sinottica:

- Unisono L Armonica fondamentale (es. do, ut).
- Quinta 2/3 L (es. do – sol).
- Quarta 3/4 L (es. do – fa).
- Tono (es. da fa a sol, ma anche da do a re) 8/9 L.
- Semitono (diatonico es. mi - fa) 243/256 L che risulta essere il fattore per passare da $64/81=(8/9)^2$ (do – re e re- mi, una terza maggiore) a 3/4 (do – fa)

---

[12] Walter Woodburn Hyde, The *Recent Discovery of an Inscribed Water-Organ at Budapest*, Transactions and Proceedings of the American Philological Association, Vol. **69**, 1938, pp. 392-410.
[13] S. Dalla Libera, *L'Organo*, p. 14-15 Milano (1956).

Esistono altre divisioni del tono che vedremo in seguito nel paragrafo dedicato ai tre generi musicali.
I Suoni Armonici sono, di fatto, delle componenti spettrali di un suono che ne determinano il particolare timbro. Sono sempre presenti nel suono reale ed hanno un peso via via minore man mano che ci si allontana dalla frequenza fondamentale.
Sul monocordo sono tutti i sottomultipli secondo numeri interi della corda vibrante.
Se la corda suona con lunghezza L (es. do1), sono presenti anche i modi L/2 (do2-ottava), L/3 (sol2-ottava+quinta), L/4 (do3 ottava+ottava), L/5 (mi3)… che corrispondono alle armoniche superiori. I suoni armonici superiori generano un accordo maggiore (es. do-mi-sol).
Nel flauto gli armonici superiori si ottengono –e si sentono con chiarezza- soffiando con forza crescente, in modo che nel tubo si instaurino 1, 2 e 3… oscillazioni complete a frequenze eguale, doppia, tripla… della fondamentale.
Esistono anche i suoni armonici inferiori, corrispondenti a 2L, 3L… misurati sperimentalmente per la prima volta da Michelangelo Abbado nel 1964[14]. Partendo da un do5 con corda lunga L, si compila la seguente tabella, in cui è presente un accordo minore.[15] Le note in grassetto sono quelle che sono multipli secondo un fattore 2, e quindi sono ottave. Poi 3L è i ¾ di 4L, quindi è un fa3, e 6L è il suo doppio, quindi un'ottava sotto, fa2. Per sapere che note sono 5L e 7L conviene anticipare il rapporto di semitono temperato che come si vede nel paragrafo dedicato al temperamento equabile è pari a 0.944. Dunque partendo dalla corda lunga 8L con il salire di 2 semitoni si ottiene $0.944^2 \approx 0.891$, mentre 7/8=0.875. Salendo di tre semitoni si ottiene un rapporto $0.944^3 \approx 0.841$, che identifica una nota più alta di quel rapporto. Dunque il suono armonico corrispondente a 7L è un comma più acuto del re2, e tre comma meno del re#2. Nel pianoforte suoniamo il re2. E allo stesso modo si verifica che 5L, che è il 5/8=0.625 del do2 corrisponde a tre semitoni sopra il fa2: $6 \cdot 0.841 \approx 5.04$, cioè il sol#2. L'accordo re fa sol# è minore diminuito, composto di due terze minori, e risolve naturalmente nell'accordo di do maggiore do mi sol.

| L | **do5** |
|---|---|
| 2L | **do4** |
| 3L | fa3 |
| 4L | **do3** |
| 5L | sol#2 |
| 6L | fa2 |
| 7L | re2 |
| 8L | **do2** |

Tavola dei suoni armonici inferiori del do5.

Per ottenere la serie degli armonici superiori si usa lo stesso metodo usato per gli inferiori.
Sono due i casi per cui occorre ricorrere al semitono temperato: L/5 ed L/7. Per L/5=0.2, abbiamo che dal do7 lungo L/4=0.25 si sale di 4 semitoni cioè si moltiplica per un fattore $0.944^4 \approx 0.794$ e si ottiene 0.198, con quattro semitoni sopra al do7 si arriva al mi7. Mentre dal sol7 L/6=0.167 si sale di 3 semitoni cioè si moltiplica per $0.944^3 \approx 0.841$ e si ottiene $0.14 \approx 1/7 \approx 0.143$, così si arriva al si bemolle 7. [16]

---

[14] M. Abbado, *Armonici Inferiori*, Enciclopedia della Musica vol. **1** p. 132 Rizzoli Ricordi, Milano (1972).
[15] A. Cavanna, *Corso Completo di Teoria Musicale*, Milano (1955) p. 20.
[16] L'accordatura del pianoforte è soggetta all'inarmonicità delle sue corde. Nelle ottave centrali le vibrazioni sono quasi armoniche, ed un bravo accordatore stabilisce le ottave ad un rapporto di frequenze 2:1 molto accuratamente. Ma nell'ottava più acuta, dove le corde sono relativamente rigide, né l'accordatore né la maggior parte degli ascoltatori saranno soddisfatti se le ottave (come do7-do8) non saranno disposte con rapporti più ampi come 2.025:1. Un effetto simile esiste per le note più basse. La rigidità delle corde determina armonici dissonanti rispetto alle corde ideali e perciò conviene modificarne leggermente i rapporti (D.E. Hall, *Musical Acoustics*, California, 2002 p. 197-198).

| L   | **do5** |
|-----|---------|
| L/2 | **do6** |
| L/3 | sol6    |
| L/4 | **do7** |
| L/5 | mi7     |
| L/6 | sol7    |
| L/7 | si *b*7 |
| L/8 | **do8** |

Tavola dei suoni armonici superiori del do5.
La generazione degli accordi maggiori e minori come armonici superiori e inferiori di una singola nota mostra come le leggi dell'armonia sono indissolubilmente legate ai numeri interi.

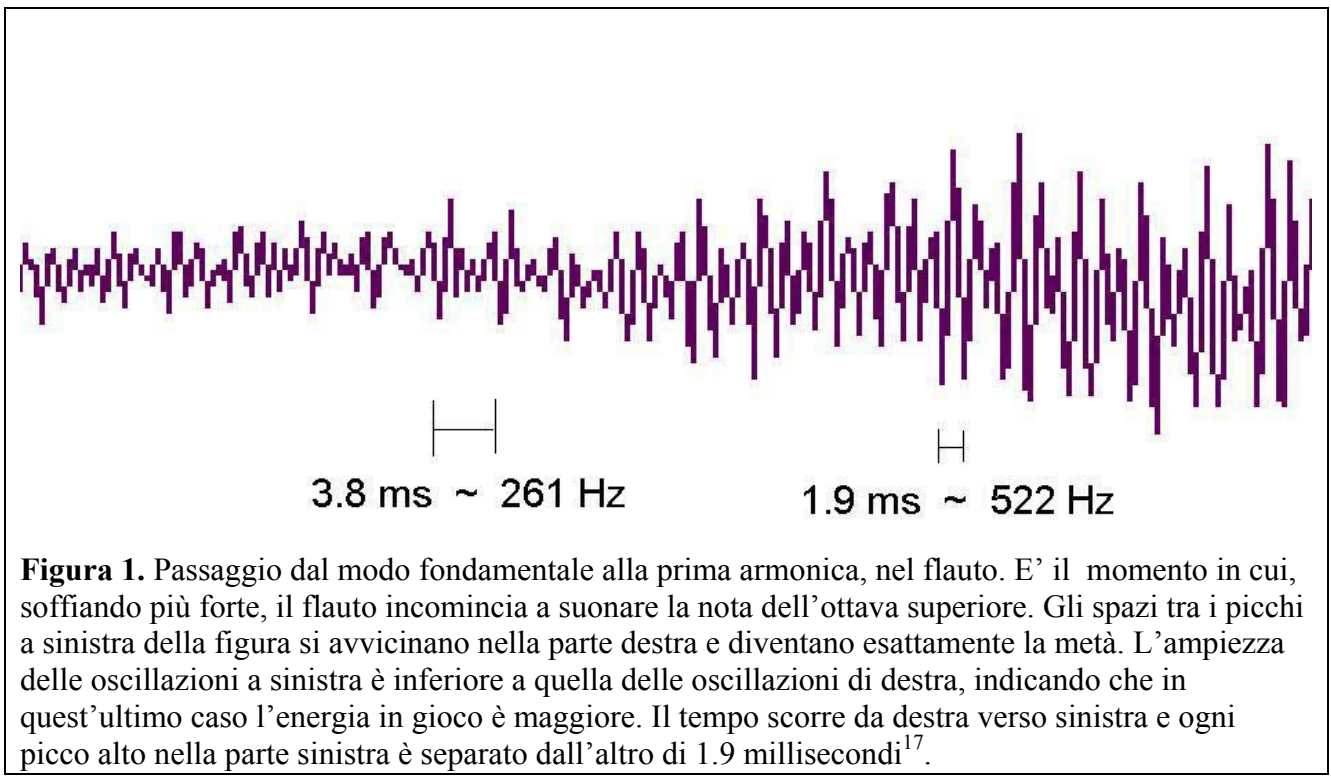

**Figura 1.** Passaggio dal modo fondamentale alla prima armonica, nel flauto. E' il momento in cui, soffiando più forte, il flauto incomincia a suonare la nota dell'ottava superiore. Gli spazi tra i picchi a sinistra della figura si avvicinano nella parte destra e diventano esattamente la metà. L'ampiezza delle oscillazioni a sinistra è inferiore a quella delle oscillazioni di destra, indicando che in quest'ultimo caso l'energia in gioco è maggiore. Il tempo scorre da destra verso sinistra e ogni picco alto nella parte sinistra è separato dall'altro di 1.9 millisecondi[17].

---

[17] L'immagine è stata ottenuta utilizzando un registratore digitale e poi visualizzando l'audio con Windows Media Player (WMP) modalità barre e onde "raggio elettrico" e stampando la schermata con il tasto speciale stamp. Si possono vedere in questo modo le sinusoidi a frequenze doppie e triple della fondamentale.
Usando la trasformata di Fourier (FFT) si può visualizzare in tempo reale lo spettro sonoro con i picchi degli armonici superiori sempre con WMP nella modalità barre e onde "tempesta di fuoco". Di fatto WMP è un vero e proprio analizzatore di segnale, anche se non fornisce unità di misura e bisogna tararlo.

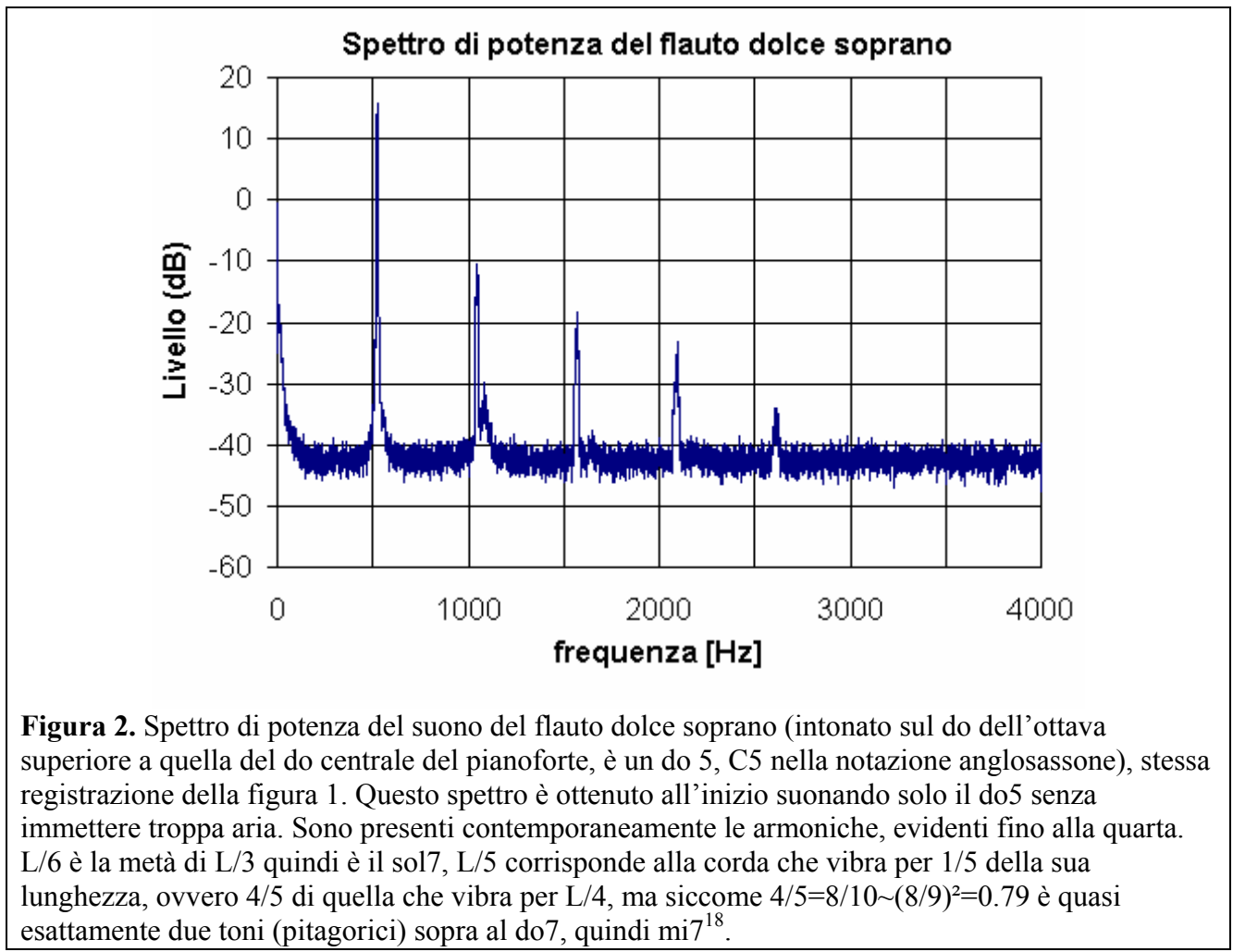

**Figura 2.** Spettro di potenza del suono del flauto dolce soprano (intonato sul do dell'ottava superiore a quella del do centrale del pianoforte, è un do 5, C5 nella notazione anglosassone), stessa registrazione della figura 1. Questo spettro è ottenuto all'inizio suonando solo il do5 senza immettere troppa aria. Sono presenti contemporaneamente le armoniche, evidenti fino alla quarta. L/6 è la metà di L/3 quindi è il sol7, L/5 corrisponde alla corda che vibra per 1/5 della sua lunghezza, ovvero 4/5 di quella che vibra per L/4, ma siccome 4/5=8/10~(8/9)²=0.79 è quasi esattamente due toni (pitagorici) sopra al do7, quindi mi7[18].

Si vede quindi che l'accordo di do maggiore, do mi sol è già naturalmente presente nella nota fondamentale, ciò ne spiega l'armonicità e la conseguente gradevolezza.
Le proporzioni della scala pitagorica e la conoscenza della frequenza di 440 Hz[19] del La sopra il do centrale del pianoforte ci permette di ricavare le frequenze del do centrale e della sua ottava superiore. Il do centrale è una quinta + un tono sotto, quindi la sua frequenza e 2/3·8/9=16/27 di 440 Hz cioè 261 Hz. Quella dell'ottava superiore è il doppio cioè 522 Hz.

## *Il temperamento musicale equabile*

La difficoltà di ricavare esattamente le note per frazioni via via più piccole della corda vibrante, quando entrano in gioco numeri primi come 5, 7, 11 … mostra la necessità pratica di avere un semitono universale, di ampiezza uniforme. Il frate francescano Gioseffo Zarlino (1517-1590) e poi Andreas Werkmeister (1645-1706) proposero quello che oggi è chiamato il temperamento equabile (1687) dividendo l'ottava in 12 semitoni uguali, ciascuno corrispondente ad una corda lunga $^{12}\sqrt{1/2}\approx 0.9439$ volte la precedente.

---

[18] L'immagine è stata ottenuta graficando lo spettro calcolato dal programma Audacity 1.3, con un campionamento del suono a 8 kHz fatto con un registratore digitale USB commerciale Creative da 512 MB.

[19] Oggi gli accordatori usano accordare il la sopra il do centrale dei pianoforti a 442 Hz. In altri paesi ed in altre epoche quel la ha avuto diverse frequenze. Il la3 del diapason è un suono base privo di timbro e dopo il congresso di Londra del 1951 ha un frequenza di 440 Hertz alla temperatura di 20 °C, ma dal 1951 questo suono base tende ancora ad innalzarsi, secondo un processo ininterrotto. Nel 1859 a Parigi il diapason era fissato a 435 Hz, ma l'intonazione di qualche organo superstite del 600-700 talvolta si trova quasi 2 toni sotto l'attuale diapason. Dal che si può dedurre che alcune note sopracute che si trovano nelle parti dei cantanti dell'Ottocento, potevano essere un secolo fa più agevolmente intonate, come riporta a p. 40L. Pinzauti, *Gli Arnesi della Musica*, Firenze (1973).

L'orecchio umano, se non educato come quello dei musicisti più dotati, non è capace di percepire le differenze tra i suoni prodotti da $0.7938\,L = (^{12}\sqrt{1/2})^4\,L$ che è il mi ottenuto salendo di 4 semitoni "equabili" dal do5, e $0.7901\,L = (8/9)^2\,L$ che è il *mi5'* ottenuto salendo di 2 toni pitagorici dal do5. E allo stesso modo il mi5 ottenuto scendendo di 2 ottave dal suono armonico mi7 (con corda L/5) *mi5''* con corda $4/5\,L = 0.80\,L$ è indistinguibile dai primi due, eccetto nel caso che vengano suonati contemporaneamente quando si verificherebbero dei battimenti, che si sentono come una lenta modulazione dell'ampiezza del suono principale.

Dunque con l'introduzione dei numeri irrazionali è stato semplificato il mondo dei semitoni e Johann Sebastian Bach (1685-1750) ha dimostrato con il suo Clavicembalo ben Temperato (48 preludi e fughe in tutte le tonalità maggiori e minori in 2 volumi: 1718-22 e 1738-42) la bontà di questa soluzione anche sotto il profilo dell'armonia.

## *Semitoni diatonici e cromatici e i tre generi musicali*

Al tempo di Gerberto, era ancora in uso la divisione della musica in tre generi: diatonico, cromatico ed enarmonico, di cui oggi si conserva ancora la memoria nei nomi dei semitoni diatonici e cromatici.

Il temperamento equabile ha eliminato, da un punto di vista quantitativo, e per gli strumenti a tastiera come il pianoforte, la differenza di ampiezza tra i semitoni nei diversi generi musicali. Il semitono diatonico oggi viene spiegato come quel semitono a cui corrisponde un cambio di nome della nota, ad esempio gli intervalli mi-fa oppure si-do, oppure anche do-re bemolle etc.. Nel semitono cromatico si conserva il nome: si-si bemolle, do-do diesis[20].

Nel passato il bemolle era un segno usato solo per il si (B nella denominazione antica ed in uso nei paesi anglosassoni), ed era il b rotundum, il b quadratum divenne poi l'attuale bequadro che annulla le alterazioni, ed il diesis # ne è un'ulteriore variante.

Il tono era suddiviso in 9 comma, ed il semitono diatonico ne copriva 4, mentre quello cromatico 5,[21] con il temperamento equabile ne spettano 4 ½ ad entrambi.

L'origine della differenza di ampiezza dei vari semitoni era nei tre generi musicali, come descritti da Aristossèno.

K. J. Sachs riporta le seguenti tabelle esplicative che aiutano a comprendere le differenze tra i tre generi esaminando le possibili suddivisioni del primo intervallo di quarta (discendente):

| Intervallo/Genere | Diatonico | Cromatico | Enarmonico |
|---|---|---|---|
| Tono t o ditono d | 9:8=1.125=t | 19:16=1.1875=$t_1$ | 81:64=1.2656=d |
| Tono $t_1$ o quarto di tono q | 9:8=1.125=t<br>t·t=t²=1.2656 | 81:76=1.0658 =$t_2$<br>$t_2\cdot t_1$=1.2656 | 499:486=1.0267 =q<br>q·d=1.299 |
| Semitono s o quarto di tono q | 256:243=1.0535=s<br>t²s=1.333 | 256:243=1.0535=s<br>$t_2\cdot t_1\cdot s$=1.333 | 512:499=1.026 =q<br>q²·d=1.333 |

Ho riportato in questa tabella anche i valori decimali ed i risultati parziali ad ogni intervallo, ad esempio dopo il primo tono $t_1$, dopo il primo ed il secondo tono $t_2\cdot t_1$ e dopo i primi due toni ed il semitono $t_2\cdot t_1\cdot s$. Si notino gli intervalli diversi, e nell'enarmonico c'è il ditono e i due quarti di tono. Segue la tabella con le scale nei tre generi. La scala discendente copre un intervallo di settima, e poi ripete l'ultima quarta fino all'ottava inferiore.

---

[20] Così p. es. in Adolfo Cavanna, *Corso Completo di Teoria Musicale*, Milano 1955, p. 17.
[21] A. Cavanna, *op. cit.*, p. 18. Il do # è poco più alto del re *b*.

| Intervalli (diat.) | Diatonico | Cromatico | Enarmonico |
|---|---|---|---|
| Ut (La) unisono | 1152 nete | 1152 hyperbolaeon (La) | 1152   (La) |
| La (Sol) - tono | 1296 paranete | 1368           (Sol *b*) | 1458   (Sol *bb*) |
| Sol (Fa)  -tono | 1458 trite | 1458             (Fa) | 1497   (Fa*) |
| Fa (Mi) -semitono | 1536 nete | 1536 diezeugomenon (Mi) | 1536   (Mi) |
| Mi (Re) -tono | 1728 paranete | 1824           (Re *b*) | 1944   (Re *bb*) |
| Re- (Do)- tono | 1944 trite | 1944             (Do) | 1996   (Do*) |
| Ut (Si)-semitono | 2048 paramese | 2048             (Si) | 2048   (Si) |
| Ut (Re) | 1728 nete | 1728 synemmenon   (Re) | 1728   (Re) |
| La (Do) - tono | 1944 paranete | 2052             (Do*) | 2187 (Do *bb*[22]) |
| Sol (Si *b*) - tono | 2187 trite | 2187             (Si *b*) | 2245 ½ (Si *b*\*) |
| Fa (La)-semitono | 2304 mese | 2304             (La) | 2304   La |

Nella tabella sono state rappresentate due scale discendenti: un esacordo composto da due tetracordi dorici[23], con una corda in comune, e corrisponde all'attuale scala di la minore naturale. Poi segue un tetracordo dorico: unisono, semitono, tono, tono.  I nomi delle note devono cambiare da una nota all'altra ed è stato introdotto il simbolo * come abbassamento di ¼ di tono, per il genere enarmonico.

Nella colonna degli intervalli della scala discendente diatonica ho riportato le note secondo la solmisazione di Guido d'Arezzo (ogni scala di sei note -esacordo-incominciava sempre da Ut, pur essendo diversa l'intonazione assoluta), tra parentesi ho indicato la nota corrispondente nell'attuale sistema musicale basato sulle sette note (dopo l'introduzione del si nel XVI secolo).

Il numero 2304 è il numero intero a cui convergono i tre generi musicali diatonico, cromatico ed enarmonico. Gerberto lo usa come punto di partenza della sua serie di lunghezze delle *fistulae*.

Il 2304, come altri già menzionati: 1, 2, 3, 4, 9=3²… ed altri ancora 192, 324,… che vedremo di seguito nelle scale, sono ottenuti a partire dalle potenze del 2 e del 3.

In particolare tutta la musica pitagorica era basata su relazioni armoniche che comprendevano tutti e solo i primi quattro numeri naturali, la cui somma è 10, numero già considerato sacro a Delfi ed esprimibile nella forma perfetta della divina *tetractis*.

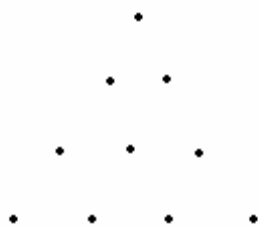

Si ricordi che i Greci rappresentavano l'unità con un punto.

---

[22] Ho anche scritto Do *bb* per non scrivere Si *b*, e coniato il Si *b*\* diminuito di ¼ di tono, per l'ultimo tetracordo enarmonico. Queste scelte sono certamente opinabili, ma riguardano solo la nomenclatura, le proporzioni matematiche determinano l'intonazione.

[23] Dal genere diatonico derivarono i generi cromatico ed enarmonico. L'accordatura cromatica si otteneva abbassando di un semitono il grado immediatamente superiore al semitono più grave. L'enarmonico invece si otteneva abbassando di una seconda maggiore il grado immediatamente superiore del semitono e di ¼ di tono il suono superiore al semitono stesso. Esempio modo dorico:
diatonico: mi,  re, do, si, la, sol, fa, mi.
cromatico: mi, re *b*,  do, si, la, sol *b*, fa mi.
enarmonico: mi, re *bb*, do*, si, la, sol *bb*, fa*, mi. Dove tra do* e si c'è ¼ di tono. Da Bruno Coltro, *Storia della Musica*, appunti.

Riassumendo la teoria pitagorica della Musica possiamo esprimere in termini matematici i seguenti concetti: la media armonica $h$ tra due numeri $a$ e $b$ è tale che $(h-a):(b-h)=a:b$.

È anche il reciproco della media aritmetica dei reciproci dei due numeri stessi $h=2/(1/a+1/b)$. Ponendo $a=L$ e $b=2L$, si ottiene che la media armonica è $4L/3$, che corrisponde ad un intervallo di quinta ascendente da $2L$ a $4/3L$ (do1-sol1) oppure di quarta discendente da $L$ a $4/3\,L$ (do2-sol1).

L'altro possibile rapporto armonico è quello di quinta ascendente che porta al do2: fa1-do2, dove il do2 è 4/3 della lunghezza del fa1. Ma do2 è già ½ del do1, cioè $L$, quindi la lunghezza del fa1 risolve l'equazione di primo grado $4/3\,L' = \frac{1}{2}\,2L=L$, da cui $L'= \frac{3}{4}\,L$.

Quindi se la nota di partenza è lunga $L$, con ¾ $L$ si ha un intervallo di quarta, con $2/3\,L$ se ne ha uno di quinta. Per salire un tono dalla quarta giusta alla quinta giusta si passa da ¾ a 2/3 moltiplicando ¾ per 8/9, dunque un tono corrisponde a $8/9\,L$. Applicando 2 toni all'unisono $L$ si arriva alla terza maggiore (do1-mi1) pari a $64/81\,L$, e i due semitoni diatonici mi-fa e si-do sono entrambi pari ad una moltiplicazione di $L$ per $243/256 \sim 0.9492\cdot L$. Infatti questa frazione risolve l'equazione $64/81\cdot L\cdot x=2/3$.

Dalla definizione di semitono cromatico, che è il complementare a quello diatonico per completare il salto di un tono si risolve l'equazione $243/256\cdot x= 8/9$ ed $x= 2048/2187 \sim 0.9364\cdot L$.

Se dividiamo il tono in 9 comma uguali, ognuno di essi corrisponderà ad una riduzione percentuale della corda iniziale lunga $L$ tale che $L\cdot x^9=8/9\cdot L$, da cui $x\approx 0.987$.

Dalle potenze $x^2\approx 0.974$ ricaviamo il valore del *quarto di tono* enarmonico, $x^4\approx 0.949$ ci dà il semitono diatonico, ed $x^5\approx 0.937$ il semitono cromatico.

Si noti che il *quarto di tono* enarmonico è ottenuto dalla media geometrica tra 243 e 256; risolvendo la proporzione $243:x=x:256$ e moltiplicando per 2 i risultati per ottenere numeri più possibile vicini agli interi: 486; 499 e 512, quindi i quarti di tono sono $486/499\cdot L \approx 499/512\cdot L\approx 0.9746\cdot L$.

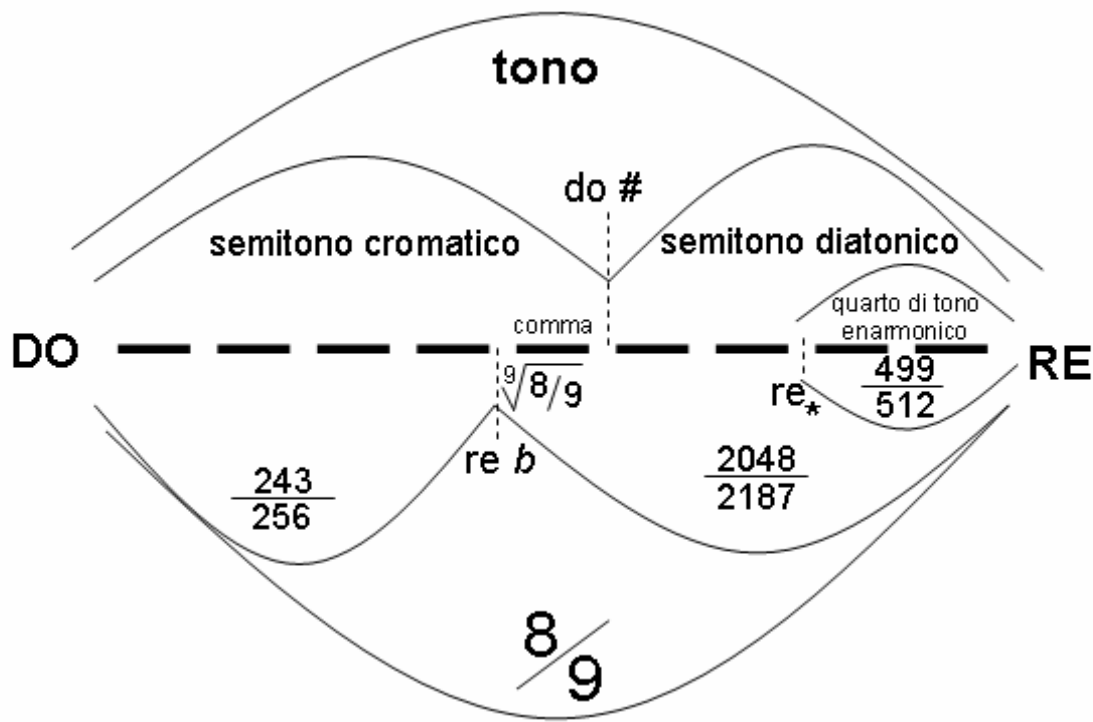

**Figura 4.** Suddivisione dell'intervallo di un tono. Il re* enarmonico è stato associato al do **x** (doppio diesis). Così come il re *bb* è stato posto in corrispondenza del suono che sia un quarto di tono enarmonico più acuto del do. Oggi i semitoni della scala temperata sono tutti eguali, ed i doppi bemolli e doppi diesis corrispondono ad un tono intero, ma il loro uso è regolato dalle leggi della tonalità. Tuttavia nella musica antica, dove si apprezzavano le sottodivisioni del tono questi intervalli erano diversi e qui se ne propone una ricostruzione con le corrispondenti frazioni, descritte nel testo.

Più complessa ancora è la suddivisione del tono secondo Aristossèno, come risulta dalla figura seguente.

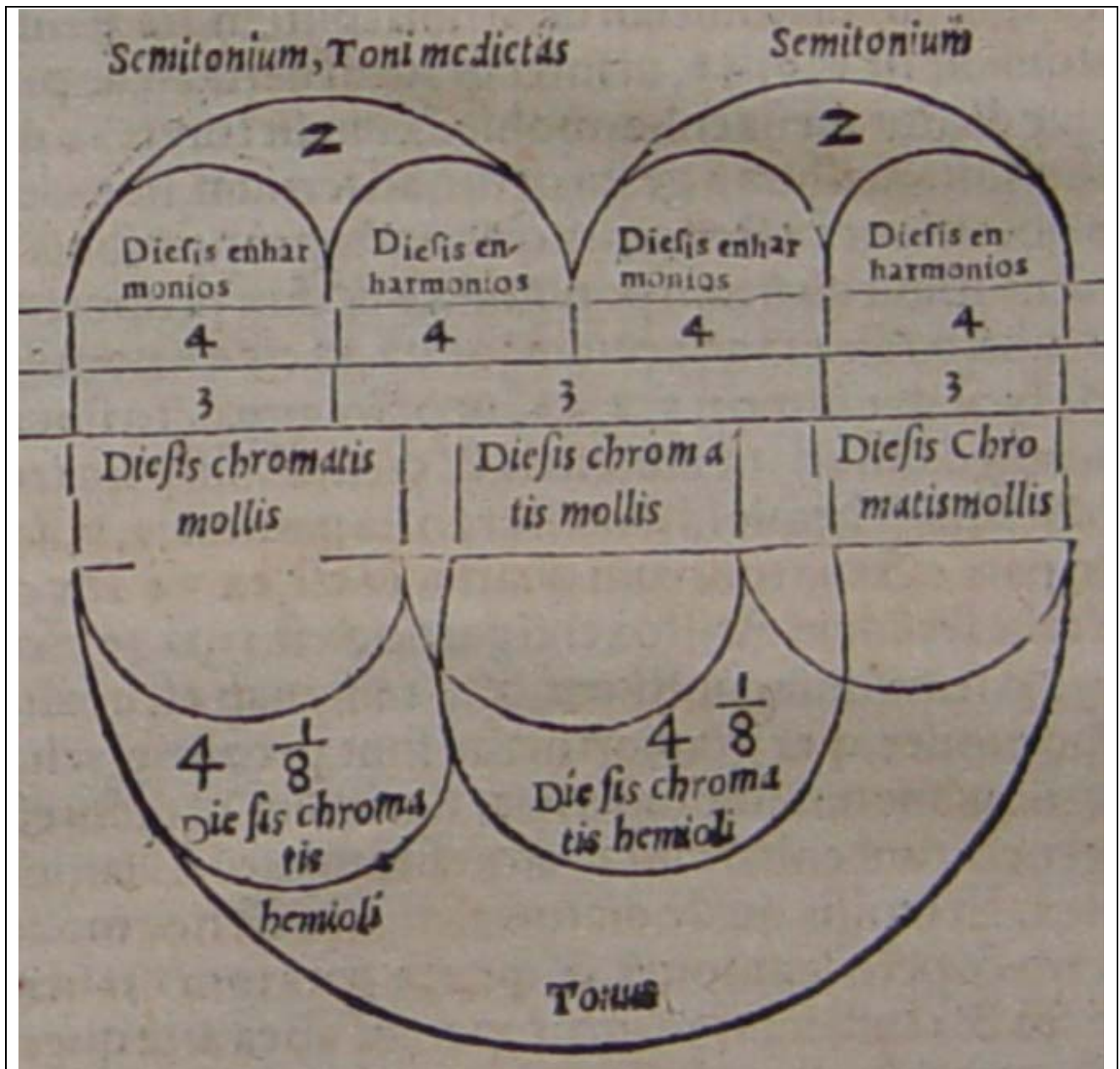

**Figura 5.** Tutti gli intervalli in cui è suddiviso il tono secondo Aristossèno.

## *Le scale*

### Scala di Calcidio

Tornando ai rapporti pitagorici la scala di Calcidio è riportata nella tabella seguente. Nella colonna di sinistra c'è il fattore per cui deve essere moltiplicata la lunghezza della corda fondamentale (es. do x 1)[24].

---
[24] La nomenclatura musicale qui usata è di poco posteriore a Gerberto: è di Guido d'Arezzo (1033) dall'inno a San Giovanni "Ut queant laxis" ed è quella più comprensibile ad un pubblico italiano con le sette note do re mi fa sol la si.

Le note diventano più acute dall'alto verso il basso. Nella colonna di destra sono i rapporti espressi con numeri decimali anziché con frazioni e sono presenti i numeri interi (tranne il primo si bemolle) che riproducono quei rapporti con il numero 324 corrispondente al do.

| La x 256/243 | 1.1852 | 384 |
| --- | --- | --- |
| Si b x 9/8 | 1.125 | 364 ½ |
| Do x 1 | 1 | 324 |
| Re x 8/9 | 0.8888 | 288 |
| Mi x 8/9 | 0.7901 | 256 |
| Fa x 243/256 | 0.75 | 243 |
| Sol x 8/9 | 0.6666 | 216 |
| La x 8/9 | 0.5926 | 192 |
| Si b x 243/256 | 0.5625 | |
| Do x 8/9 | 0.5 | |

La tabella riprodotta in figura 4 è di Michel Huglo e serve ad ottenere i numeri usati da Calcidio a partire dalle serie di potenze del 2 e del 3.

Il triangolo è ottenuto ponendo le potenze di 2 sul lato sinistro ($2^0$, $2^1$, $2^2$, $2^3$…) e le potenze intere di 3 sul lato destro con la prima in comune ($3^0=1$, $3^1$, $3^2$, $3^3$…), dopo di che ogni coppia di numeri sul lato sinistro si sommano e si completa la casella come nella figura 3.

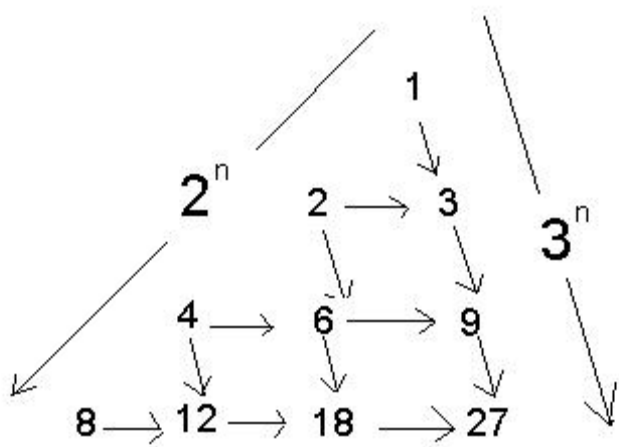

**Figura 3.** Algoritmo per realizzare il triangolo di Calcidio. Ogni casella dove convergono due frecce è la somma dei valori nelle caselle di provenienza.

Come si vede le potenze intere del 3 sono anch'esse il risultato del medesimo algoritmo di somma, indicato in figura dalle frecce convergenti nella casella dove è scritta la somma.

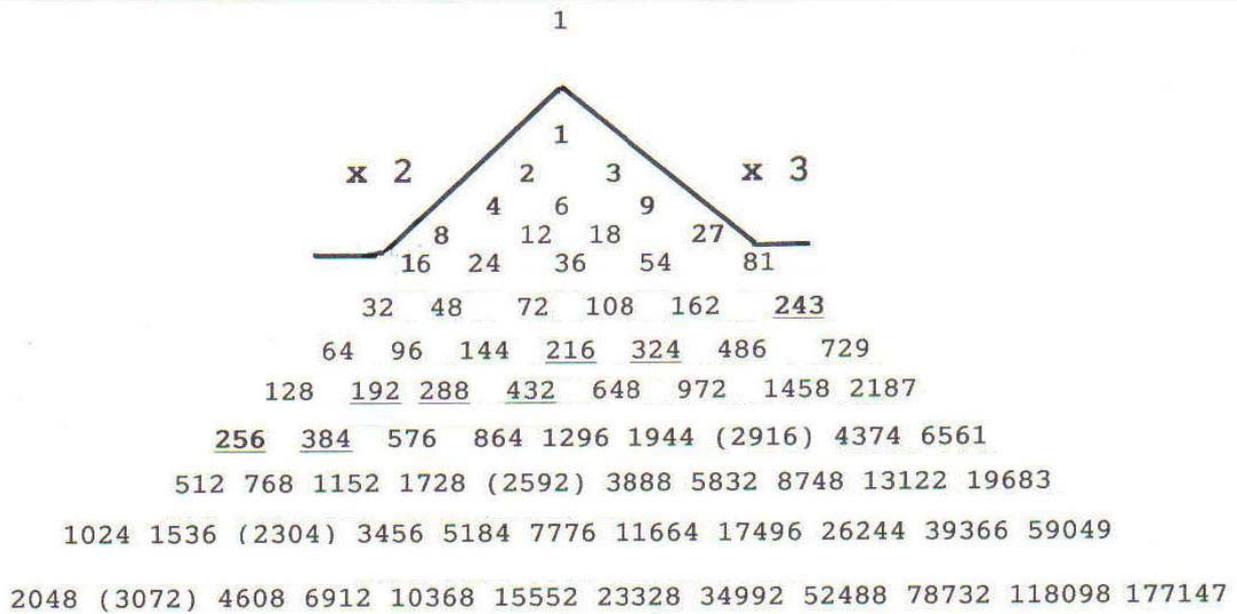

**Figura 4.** Tabella di Michel Huglo sulla derivazione dei numeri musicali di Calcidio e Boezio dalle serie delle potenze intere di 2 e di 3. Tutti i numeri della scala diatonica da 1152 a 2304 sono presenti, compreso il 2187=$3^7$ che viene dalla divisione del tono in semitono diatonico ed il suo complementare: il semitono cromatico.

## Scala di Gerberto

La scala adottata da Gerberto nella *Mensura Fistularum*, di fatto, è la scala minore naturale. Con la stessa notazione della precedente scala sono riportate le proporzioni, graficate in figura 5.

| | | |
|---|---|---|
| La x 9/8 | 1.1852 | 384 |
| Si x 256/243 | 1.0535 | 341 1/3 |
| Do x 1 | 1 | 324 |
| Re x 8/9 | 0.8888 | 288 |
| Mi x 8/9 | 0.7901 | 256 |
| Fa x 243/256 | 0.75 | 243 |
| Sol x 8/9 | 0.6666 | 216 |
| La x 8/9 | 0.5926 | 192 |
| Si x 8/9 | 0.5267 | |
| Do x 243/256 | 0.5 | |

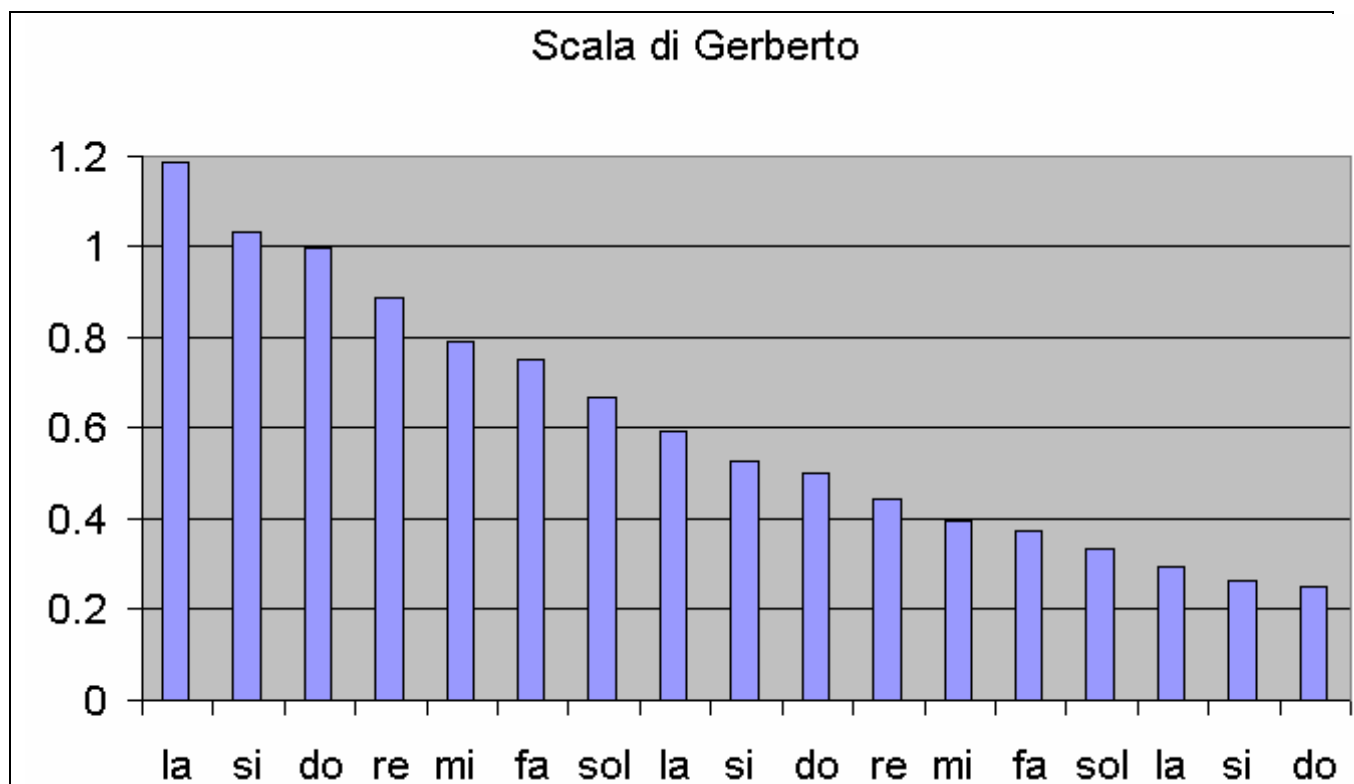

**Figura 5.** Traducendo in grafico la lunghezza delle corde per la scala di Gerberto otteniamo la visualizzazione dell'equazione di un'Arpa a corde di uguale densità lineare e uguale tensione.
I dati usati in figura 5 e nella tabella 1 e 2 sono presi da C. Meyer (1997).

## End Correction per le canne d'organo

A differenza di una corda vibrante, dove i punti di appoggio della corda determinano una regione di transizione tra il movimento e la posizione fissa molto ridotta, nelle canne d'organo la zona di transizione tra il tubo (*fistula* in latino) e l'aria esterna è tanto più grande quanto maggiore è il diametro del tubo aperto da una, o da entrambe le parti.

Quando il tubo è aperto da entrambe le parti si parla sia di correzione di bocca, mouth end correction, che di open end correction.

Dell'argomento si sono interessati molti fisici, tra i quali spiccano Daniel Bernoulli (1764) e Lord Raileigh (1870, 1926).

In una canna cilindrica di raggio r e lunga L risuona in modo stazionario un'onda sonora di lunghezza

$\lambda = L + 0.6 \cdot r$

Se la canna ha una apertura *flanged* ovvero a tromba (Raileigh, 1926) la correzione vale
$\lambda = L + 0.85 \cdot r$

Se vogliamo passare dalle corde pitagoriche alle corrispondenti lunghezze delle canne d'organo occorre applicare queste correzioni alla lunghezza per non incorrere in dissonanze.
Le canne tanto più sono sottili, quanto più vicine sono al limite pitagorico.
La "*end correction*" è valida per basse frequenze ed indipendente dalla frequenza del suono (nota suonata).

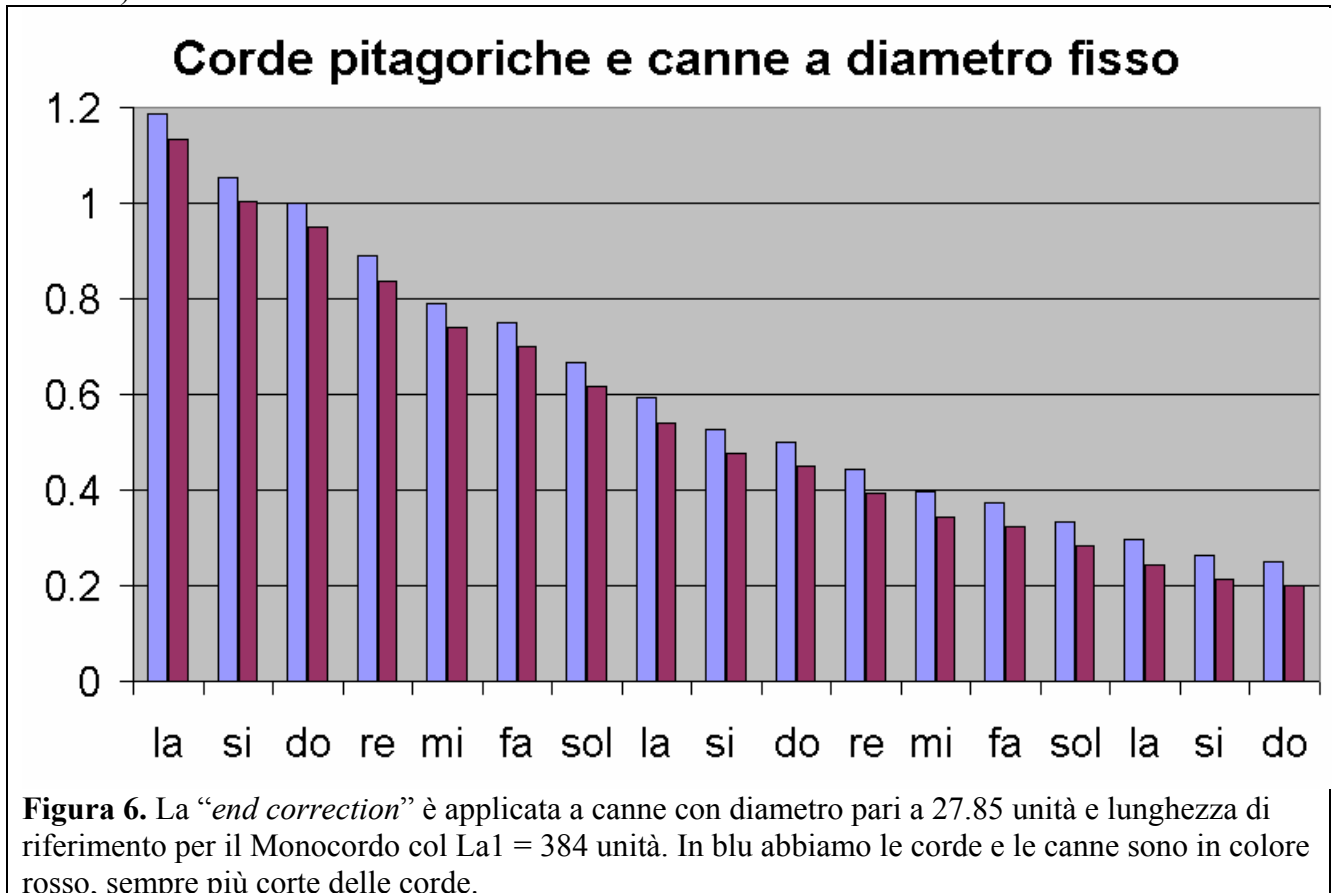

**Figura 6.** La "*end correction*" è applicata a canne con diametro pari a 27.85 unità e lunghezza di riferimento per il Monocordo col La1 = 384 unità. In blu abbiamo le corde e le canne sono in colore rosso, sempre più corte delle corde.

## *Data Igitur*

Dal Manoscritto della Biblioteca Nazionale di Madrid n. 9088 del XII secolo, intitolato Gerbertus De Commensuralitate Fistularum et Monocordi cur Non Conveniant Rogatus a Pluribus…

È l'attestazione più antica che il testo è di Gerberto, mentre Martin Gerbert nel primo volume sugli *Scriptores ecclesiastici de Musica sacra* (1784) aveva attribuito questo testo a Bernellino che era posteriore a Gerberto.

**57** Data igitur primae vel minori fistulae qualibet longitudine, sed melius videtur diametro foraminis octies longitudini dato, ubicumque deinceps tonus est faciendus, maior habeat minorem totam et diametrum et diametri eius octavam.
**58** In diatessaron maior minorem et eius tertiam et insuper diametri tertiam.
**59** In diapente maior minorem et eius mediam insuper et diametri mediam.
**60** In diapason maior minorem duplo et totum insuper diametrum.
**61** Ceterae vero facile per numeros occurrent.
**62** Est autem diametrum vel circuli, qui est in foramine ex ampliori parte, medietas vel foraminis transversitas ex deductiori parte, qua inspiratur fistula et cui foramini subiacet uva.
**63** Ut autem minoribus id ostendatur numeris, sed tantum interruptis secundum symphonias, ita erit figurandum:

    16
        9
    25
        9
    34
        19 1/8
    53 1/8
        19 1/8
    72 1/4

**64** Hi numeri interruptim positi sub exemplo consonantias tantum monstrant: 16 namque ad 34 primum diapason, 34 ad 72 et quadrantem secundum diapason, 25 in medio primi diapason ad alterum diapente, ad alterum diatessaron est, **65** sicut et 53 et octavum in medio secundi diapason ad alterum diatessaron, ad alterum diapente est; 9 et 9 et 191/8 et 191/8 sunt differentiae; **66** sed si positos hos numeros multotiens duxeris, qui interponendi sunt mox integri occurrent secundum regulam infra demonstratam.

Traduzione italiana:

**57** Data dunque una qualsiasi lunghezza alla prima canna, cioè la minore, anche se è migliore che si abbia un diametro del foro pari ad 1/8 della lunghezza, da questa inizialmente si fissa il tono, quella più grande è pari a tutta la minore aumentata del diametro e di 1/8 dello stesso.
**58** La canna [che suona la nota] di una quarta più bassa è aumentata di una terza parte della minore e di un terzo del diametro.
**59** La canna della quinta più bassa è aumentata della metà della minore e di metà del diametro.
**60** La canna dell'ottava [più grave] è il doppio della minore più un intero diametro.
**61** Le altre facilmente si ricavano dai numeri.

**62** D'altronde il diametro è del cerchio, che è nel foro dalla parte più grande, la metà ossia la larghezza del foro dalla parte più sottile, dove la canna prende aria e sotto la cui apertura sta l'ancia [uva significa ugola].
**63** Come d'altronde dai numeri minori ciò discenda, ma interrotti soltanto secondo gli accordi, così si rappresenterà:

  16
    9
  25
    9
  34
    19 1/8
  53 1/8
    19 1/8
  72 1/4

**64** Questi numeri posti interrottamente come esempio mostrano solo le consonanze: infatti 16 [sta] a 34 per la prima ottava, 34 [sta] a 72 e ¼ per la seconda ottava, 25 sta come medio tra la prima ottava l'altra quinta all'altra quarta, **65** e così 53 e 1/8 sta come medio della seconda ottava all'altra quarta e all'altra quinta ; 9 e 9 e 19 1/8 e 19 1/8 sono differenze; **66** ma se condurrai più volte questi numeri posti, che sono da interporre subito interi occorrono secondo la regola dimostrata più avanti.

Passando ad una traduzione in formule abbiamo:

**57** La nota di partenza è fatta da una canna lunga 8D; un tono più basso si ottiene con
$(9 + 1/8)D = 73/8\ D$
**58** La quarta più bassa con $8 \cdot 4/3 \cdot D + D/3 = 33/3 \cdot D = 11 \cdot D$
**59** La quinta più bassa con $8 \cdot 3/2 \cdot D + D/2 = 25/2 \cdot D$
**60** L'ottava più bassa con $8 \cdot 2D + D = 17D$
se poniamo D=2 otteniamo la tabella

| | |
|---|---|
| Nota di partenza | 16 |
| Tono sotto | 18 ¼ |
| Quarta sotto | 22 |
| Quinta sotto | 25 |
| Ottava sotto | 34 |

**64** Vale la proporzione 16:34=34:72 ¼ , il numero 25 è la media aritmetica tra 16 e 34. **65** così come 53 e 1/8 è medio tra 34 e 72 ¼.

La correzione tipo "*end correction*" che minimizza[25] gli scarti al quadrato tra i dati tabulati da Gerberto e quelli ottenuti applicando correzioni del tipo α·r vale α=2.

Vediamo il procedimento a partire da una canna lunga 16 unità, di diametro 2 unità e la "corda equivalente" di tensione opportuna e che suoni la stessa nota..La corda è sempre più lunga della fistola, e passando da un tono all'altro segue le proporzioni pitagoriche.
Le operazioni algebriche sono mostrate nel seguente ordine: la prima è la correzione gerbertiana L+α·r con α=2, per ottenere la lunghezza della corda equivalente ad una data fistula; la seconda mostra come la misura della lunghezza della corda equivalente sia nelle dovute proporzioni pitagoriche con le altre (1; 9/8; 4/3; 3/2).

---

[25] Precisamente α=2 annulla gli scarti al quadrato. Ci si poteva aspettare che i dati empirici non seguissero esattamente una legge L'=L+α·r, ed invece è venuto un accordo perfetto. Tuttavia vedremo per la seconda ottava che questa legge non viene più seguita.

| Intervallo | Corda equivalente | *fistula* |
|---|---|---|
| Unisono | 18=16+1·2=18·1 | 16 |
| Tono | 20¼=18 ¼+1·2=81/4=18·9/8 | 18 ¼ |
| Diatesseron | 24=22+1·2=18·4/3 | 22 |
| Diapente | 27=25+1·2=18·3/2 | 25 |
| Diapason | 36=34+1·2=18·2 | 34 |

A questo punto per passare all'ottava inferiore si dovrebbe applicare di nuovo lo stesso algoritmo che porterebbe da un diapason all'altro a 70 unità di lunghezza, infatti usando la formula del n. 60 abbiamo che l'ottava inferiore è il doppio (34·2=68) più la correzione pari al diametro (2), invece nella tabella di Gerberto abbiamo una lunghezza di 72 ¼ unità.

Gerberto ottiene questo valore applicando direttamente la proporzione
16:34=34: x ed x= 34²/16= 72 + ¼ .

In questo calcolo Gerberto non usa più le correzioni empiriche valutate pari ad un intero diametro, per ottenere la vera lunghezza della fistula a partire dalla lunghezza equivalente (pitagorica).

Se consideriamo un altro algoritmo, basato sulle differenze, vediamo che dalla nota fondamentale di 16 unità alla ottava successiva lunga 34 unità c'è una differenza di 18=(13+½ )·2/3·2, mentre tra 34 e 72 ¼ c'è una differenza di 38 ¼ = (14+1/3+1/144+1/288 )·2/3·2.
Dunque Gerberto usava l'algoritmo moltiplicativo –basato sulle differenze per la prima ottava- per passare da un'ottava a quella più grave, anziché quello iterativo (L'=L+α·r) da applicare ogni volta alla nota successiva L' a partire dalla lunghezza L della precedente.
Ma il moltiplicatore cambia, da 13.5 a 14.34375=(14+1/3+1/144+1/288).
Questo non è un modo per implementare la variazione della costante di *end correction* con la frequenza della nota. Passando ad ottave più basse si va verso frequenze più basse, ed è proprio nel limite di basse frequenze per le quali vale la legge α=0.6 teorizzata da Lord Rayleigh.

Altri manoscritti sullo stesso argomento collazionati da K. J. Sachs e posti sul web da C. Meyer[26] riportano le stesse proporzioni, ma quello di Monaco (Bayerische Staatsbibliotechek (D-Mbs), Clm 23577, f.78v (M12) applica alle *fistulae* le stesse proporzioni pitagoriche, senza alcuna *end correction* come si vede chiaramente dall'ottava *fistula* che è semplicemente il doppio della prima:

**1** Prima fistula in octo dividitur.
**2** Secunda habet primam in se et eius octavam partem.
**3** Tertia secundam et eius octavam partem.
**4** Quarta fistula habet in se primam et eius tertiam partem.
**5** Quinta primam in se et eius medietatem.
**6** Sexta quintam in se et eius octavam partem.
**7** Septima fistula habet quartam in se et eius tertiam partem.
**8** Octava fistula habet primam in se duplam.

Nel manoscritto di Oxford "Circa Latitudinem" è considerata la proporzione con diametro (latitudo) pari ad 1/7 della lunghezza (longitudo). Più avanti è presentata la proporzione 22:7 che era l'antica approssimazione per il π=3.14, rapporto tra circonferenza e diametro.

OXFORD, Bodleian Library (GB-Ob), Bodley 300 (S.C. 2474), 119va-b (O)

---

[26] C. Meyer nel *Lexicon musicm Latinum* su http://www.lml.badw.de/info/fist01.htm

PRINCETON (NJ), University Library (US-PRu), MS Garrett 95, 85v-86r (p. 176-177) (Pr)

**1** Circa latitudinem fistularum est sciendum, quod quaelibet fistularum prima continebit in latitudine septimam partem longitudinis suae, si sit de stagno; et si sit de plumbo, continebit quintam partem, et hoc rariter verum est, fallit tamen in quibusdam locis, ut in inferioribus fistulis semper patet.

**2** Cape ergo longitudinem primi .C. et divide eam in septem partes aequales, et illa septima pars erit latitudo sua.

**9** Nota tamen, quod os fistulae continebit quartam partem diametri sui, diametrum vero tertia pars fere latitudinis fistulae; fere dico, quia aliquid debet minui de tertia parte, nam circulus continet quantitatem diametri ter et septimam partem tertiae partis vel ipsius diametri, unde proportio circuli ad ipsum diametrum est sicut proportio 22 ad 7, quia 22 continet 7 ter et unitatem, quae est pars septima eiusdem ternarii.

Nel manoscritto che inizia con le parole "Fac tibi" è riportata la dimensione, seppure approssimata, della prima fistula:
una ulna e mezzo, cioè tra 35 e 45 cm[27].
W. T. Atcherson (1983) nota come la maggior parte dei trattati non diano la lunghezza della prima canna, comprovando che si tratta di trattati di musica teorica più che pratica. Le formule in questione sono più un'estensione di numeri "Pseudo-Pitagorici" che un reale tentativo di avvicinamento tra teoria e pratica.
Anche P. Williams, recensendo il secondo dei volumi di K. J. Sachs sulle canne d'organo, afferma che i teorici, autori di questi testi, non erano pratici per niente, almeno nel senso di dare istruzioni per un potenziale costruttore di organi, o altro tipo di musicista. La cultura dei manoscritti era diversa da quella degli attuali manuali nell'era tecnologica.
La precedente citazione del pi greco nel manoscritto che inizia con "Circa latitudinem" sembra confermare il fine didattico più che pratico di queste nozioni ed algoritmi.

Olim ST. BLASIEN, Stiftsbibliothek, Côté inconnue (manoscritto distrutto nell'incendio del 23 luglio 1768 dove l'abate Martin Gerbert vide andare in fumo la sua prima stesura dei suoi *Scriptores Ecclesiastici de Musica Sacra* e gli occorreranno 11 anni per ricostituire la sua pubblicazione a St. Blaise nel 1779[28]).

**1** Fac tibi fistulam secundum aestimationem, utpote unius ulnae et dimidiae longam.
**2** Huius latitudinem divide in octo partes, et octavam concede extra mensuram; quod reliquum est usque ad plectrum id est linguam divide in novem partes, et ex illis da octo secundae fistulae; haec erit longitudo eius a plectro sursum, et habes tonum inter duas fistulas.

---

[27] Stime riferite rispettivamente ad 1.50m e 1.90m di statura.
[28] M. Huglo. *Gerberto Teorico Musicale, visto dall'anno 2000*, Bobbio 2001, p. 220-221.

## *Correzione gerbertiana e forma della canna d'organo*

Tornando al testo di Gerberto, *Data Igitur*, l'accordo tra dati tabulati e canne cilindriche aperte con una end correction r·α con α=2 è differente dalla *end correction* teorica di α=0.6 potrebbe indicare che Gerberto considerava anche l'influenza della *correzione di bocca*, che può incidere a seconda della geometria in gioco, per un fattore anche molto superiore alla *end correction*.

Per misurare la correzione acustica totale di un tubo sonoro occorre misurare prima la sola correzione di bocca, tappando l'apertura libera.
In questo modo il tubo risuona a λ/4, e $\lambda_1/4$ è proprio la lunghezza efficace del tubo.
La differenza b=|L-$\lambda_1$/4| è la correzione di bocca.
Adopero il segno di valore assoluto perché $\lambda_1$/4 nel flauto dolce è più corta della lunghezza dello strumento, mentre nel tubo sonoro da oltre 1 m di lunghezza è più lunga del tubo.
Poi si lascia il tubo aperto, in modo che risuoni a λ/2, e $\lambda_2$/2 è la nuova lunghezza efficace del tubo.
La differenza b+e=|$\lambda_2$/2-L| è la somma delle due correzioni acustiche di bocca e di *open end*.[29]

Ecco l'esempio di una canna a sezione rettangolare, ripreso da J. J. Dammerud.

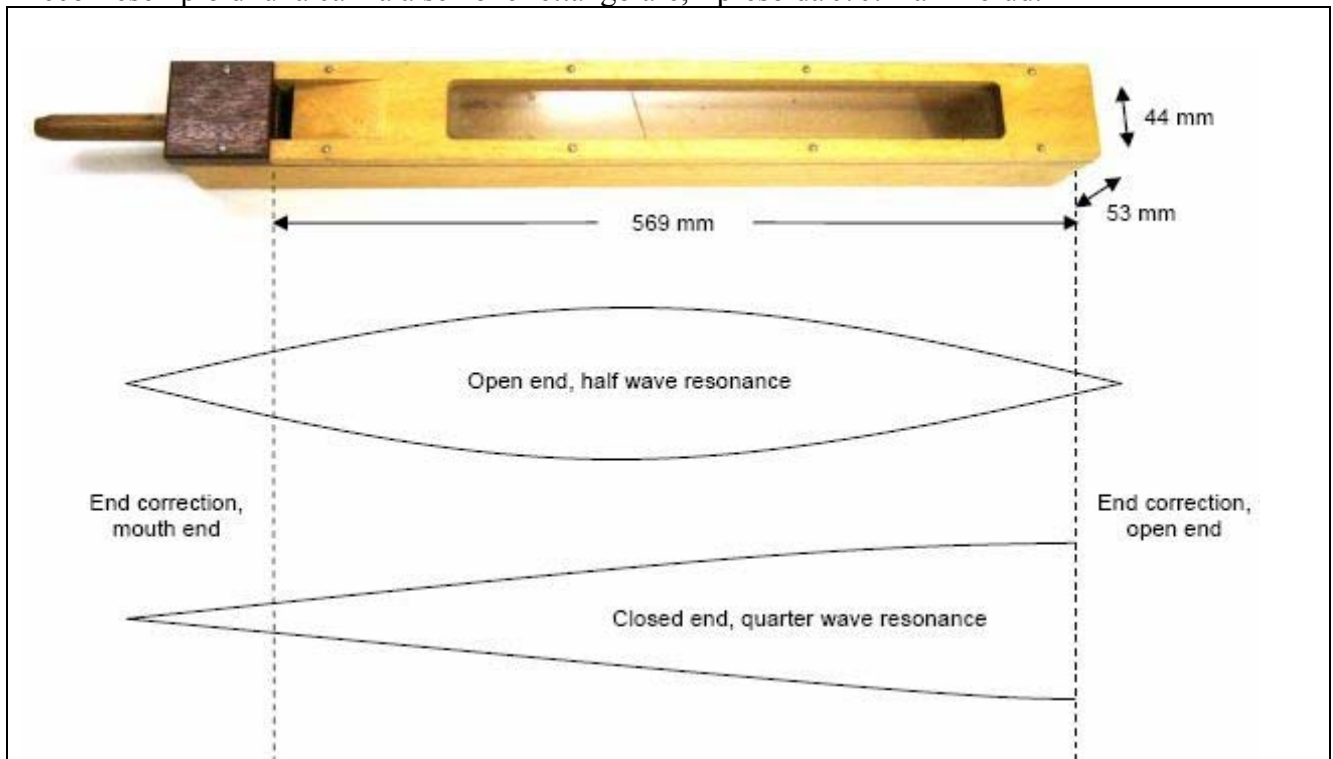

**Figura 7.** Rappresentazione[30] visiva delle lunghezze d'onda che risuonano nella canna a sezione rettangolare. La correzione di bocca risulta oltre 2 volte il diametro della massima sezione, mentre quella di apertura libera obbedisce alle predizioni teoriche di Lord Raileigh con un valore α~0.6.
La formula di end correction per tubo cilindrico aperto ricavata per via teorica è dunque ben verificata, ma è evidente che la correzione di bocca incide molto di più.

---

[29] J. J. Dammerud ha messo in rete l'applicazione di questo metodo ad una canna acustica a sezione rettangolare. Nello stesso sito ha indicato anche il software con cui ha analizzato lo spettro dei suoni, Audacity, con il quale ho potuto realizzare le misure di frequenza molto accurate delle esperienze che ho descritto in questo articolo.
[30] http://people.bath.ac.uk/jjd22 (2006)

I dati di Gerberto sono consistenti con tale osservazione, anche se l'unica valutazione che possiamo sulle *corrections* da lui usate resta quella dei minimi quadrati, che ci ha dato un valore di α=2, poiché non conosciamo la forma esatta delle canne da lui usate, nella parte dove ricevono l'aria.

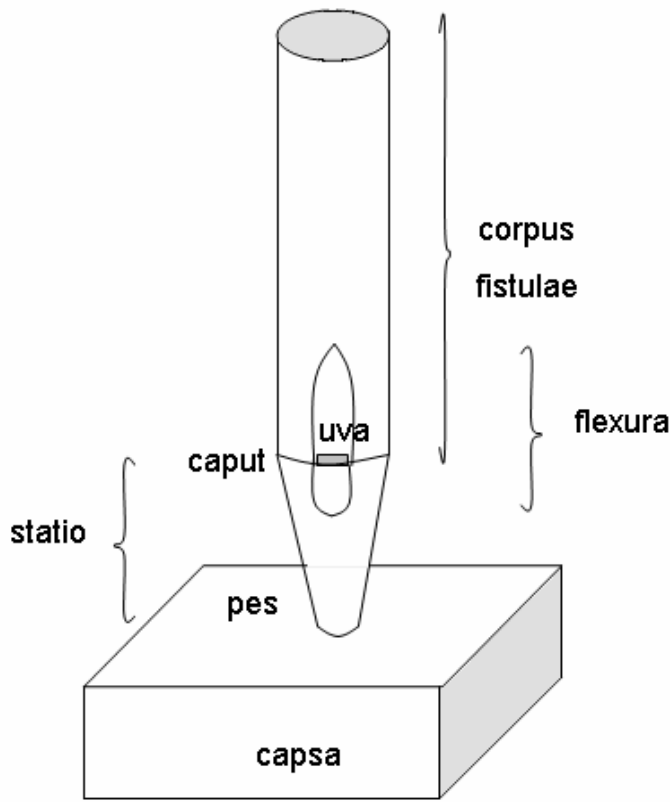

**Figura 8.** Struttura di una *fistula* (adattata da K. J. Sachs vol. 2 p. 57). La *end correction* complessiva usata da Gerberto è 3 volte quella della sola *open end*, a causa della forma dell'ancia (*uva*) e della parte (*statio*) che si inserisce nella *capsa*.

Nei paragrafi seguenti misuriamo correzioni di *end* e di bocca, conoscendo il valore preciso della velocità del suono.

## La velocità del suono

Per fare queste misure occorre tenere conto in modo esatto anche della temperatura, infatti la velocità del suono varia del 3% tra 0 °C e 20 °C.
La formula per la velocità del suono in funzione di pressione e densità[31] è

(1) $\qquad c=\sqrt{1.4(P/[Pa])/(\rho/[Kg \cdot m^{-3}])}$

Assumendo l'equazione di stato dei gas perfetti la pressione è proporzionale alla densità e alla temperatura secondo l'equazione $P \sim 288 \cdot \rho(T+273.16)$ con T misurata in °C. Per cui sostituendo

---
[31] H. C. Ohanian, *Physics*, Norton & Co., New York (1989) p. 443.

nella (1) al posto della pressione P si ottiene l'equazione della velocità del suono funzione della temperatura T [°C]

(2) $\qquad c=\sqrt{403\cdot(T+273.16)}$

Da questa si verifica che a 0°C la velocità vale 331.8 m/s mentre a 20°C diventa 343.8 m/s. La variazione è di 13 m/s su 20°C, ovvero 0.65 m/s/°C.
Notare che nell'equazione (2) la dipendenza dalla pressione e dalla densità presente nella (1) sparisce, e nella pratica è cruciale misurare la temperatura alla quale si fanno gli esperimenti, piuttosto che la pressione o la densità.[32]

La variazione di velocità del suono con la temperatura è molto importante per l'intonazione degli organi, infatti la lunghezza di una canna non può variare, e la lunghezza d'onda che ci vibra dentro vale $\lambda/2 = L + \alpha \cdot r$ dipende esclusivamente dalle dimensioni fisse della canna. La velocità del suono c è in relazione con la frequenza ν con l'equazione $c = \lambda \cdot \nu$, da cui $\nu = c/\lambda$, per cui se la temperatura aumenta di 10°C c aumenta di 6.5 m/s cioè dell'1.8%, e anche la frequenza aumenta della stessa percentuale. Se si aveva una canna intonata sul la4 a 440 Hz a 20°C questa a 30°C suonerà una frequenza più alta dell'1.8%, cioè 448 Hz, che corrisponde a poco più di un comma più acuto. Infatti un intervallo di un comma corrisponde alla riduzione della lunghezza di una corda di un fattore 0.987 (frequenze più alte di 1.3%), o all'aumento di 1/0.987=1.013 (frequenze più basse di 1.3%). Due comma sono 0.974 in riduzione o 1.026 in aumento(2.6% di riduzione o aumento). Variazioni piccole in senso assoluto, ma apprezzabili sotto forma di battimenti se ci sono altri strumenti a corda, la cui intonazione dipende dalla tensione della corda.
Nell'esempio appena fatto il battimento avrebbe una frequenza pari alla semidifferenza tra 440 e 448 Hz, cioè 4 Hz, ovvero una pulsazione dell'intensità del suono 4 volte al secondo.
Con una temperatura di soli 5°C superiore alla temperatura a cui l'organo è stato intonato avremmo 444 Hz ed un battimento a 2 Hz molto ben udibile.

Nel caso della misura della correzione di bocca e di *end* un errore dell'1% nella velocità del suono corrisponde ad un errore della stessa percentuale sulla lunghezza d'onda efficace del tubo sonoro, che su tubi di 1 m corrisponde ad 1 cm, che è proprio l'entità della correzione acustica da misurare. Dunque la misura della temperatura è essenziale per una determinazione precisa di *c*.[33]

## *Misura delle armoniche superiori e della dipendenza di α dalla frequenza*

In questo paragrafo affronteremo il problema delle canne reali, semplificandolo tuttavia ai casi di due tubi sonori.
La correzione di bocca e quella di *end* sono uguali, poiché ho usato un tubo aperto alle estremità di diametro crescente da 13.3 mm a 19 mm, e lungo 1171.5 mm nel primo caso, ed uno corto da 169 mme largo da 39 a 46 mm.
Non è contemplato qui il caso di *flexura* con *uva*.

---

[32] La velocità del suono rimane indipendente dalla pressione atmosferica fino ad un alto grado. Infatti la pressione critica, oltre la quale l'aria cessa di comportarsi come un gas perfetto vale 30 atmosfere. La velocità del suono vale v=331.4 m/s a 0°C ed ha una variazione con la temperatura di $\partial v/\partial T = 0.6$ m/s/°C. R. H. Kay, *Sound Propagation in the Atmosphere*, Encyclopaedic Dictionary of Physics, J. Tewlis ed., Pergamon Press, Oxford London New York Paris, 1962.
[33] Una misura della variazione della velocità del suono con la temperature si può fare proprio con la tecnica messa a punto in questi paragrafi: sapendo che la correzione di *end* per un tubo vale $0.6 \cdot r$ è possibile confrontare $c/\nu = \lambda/n$, dove n è l'ordine dell'armonica in esame, con la lunghezza L del tubo aperto ai due lati aumentata di due volte $0.6 \cdot r$. Il valore di *c* che minimizza gli scarti tra *c/ν* e $L + 1.2 \cdot r$ è la velocità del suono. L'esperimento richiede la massima precisione nella misura delle frequenze di risonanza, ma basta la prima armonica.

Ho fatto delle misure, con gli stessi criteri già esposti, su un flauto dolce soprano, ottenendo che la bocca si estende, in pratica, fino al termine della scanalatura superiore del becco. Ciò comporta che per una canna d'organo reale la correzione di bocca includa ragionevolmente tutta la *statio* e la *flexura* (come mostro in dettaglio nel paragrafo seguente).

La dipendenza di α dalla frequenza è stata studiata da Levine e Schwinger (1947) per un tubo cilindrico e l'andamento teorico segue la figure seguente.

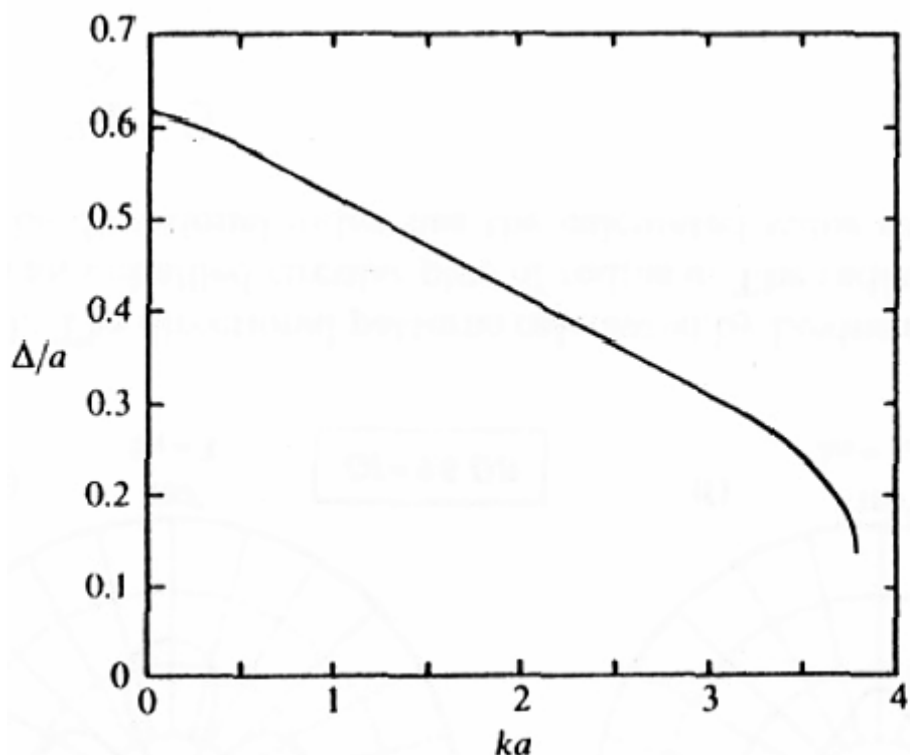

**Figura 9.** Andamento della *end correction* teorica[34] Δ normalizzata al raggio *a* del tubo cilindrico. In ascisse $k=2\pi/\lambda$ è il numero d'onda, pari a $2\pi\nu/c$, dove *c=344* m/s a T=20 °C. Il prodotto *ka* ha le dimensioni di un numero puro. Per *a*=8 mm *ka*=1 per frequenza ν=6847 Hz. Mentre per *a*=2 cm come potevano essere le *fistulae* alte un'ulna menzionate nel manoscritto andato perduto di St. Blaise, *ka*=1 per frequenza ν=2739 Hz (nota fa7, il fa più acuto nelle tastiere dei pianoforti). La differenza tra le correzioni di queste canne con *ka*=0 e *ka*=1 è di Δ=0.1·*a*=2 mm, mentre la correzione vale complessivamente 10 mm al fa7 e 12 mm alle più basse frequenze udibili.
Le canne degli organi al tempo di Gerberto, piccoli organi idraulici o organi portativi, essendo sottili, dovrebbero situarsi tutte nella regione *ka*<1, dove la correzione è quella massima α=Δ/*a*=0.6.

Seguendo il metodo per la misura delle correzioni acustiche esposto nel precedente paragrafo ho potuto misurare empiricamente la dipendenza dalla frequenza della correzione acustica sfruttando la concomitanza di varie armoniche superiori al suono fondamentale.
Innanzitutto con un diapason[35] a 440 Hz ho verificato la bontà del software Audacity[36] che in circa 10 secondi di tempo di integrazione così risolve la sua frequenza.

---

[34] Figura 8.9 del testo di Fletcher e Rossing, *The Physics of Musical Instruments*, New York (1998), fig. 8.9 ripresa da Levine e Schwinger (1948).
[35] Un ringraziamento a Lara dello *Studio 12* Pianoforti di Roma, per le registrazioni col diapason.
[36] La versione beta 1.3 è quella che consente, al momento della scrittura di questo articolo, le analisi spettrali del suono più avanzate. Il programma è freeware ed è disponibile al sito http://audacity.sourceforge.net . Trattandosi della versione Beta, è bene fare un test di taratura.

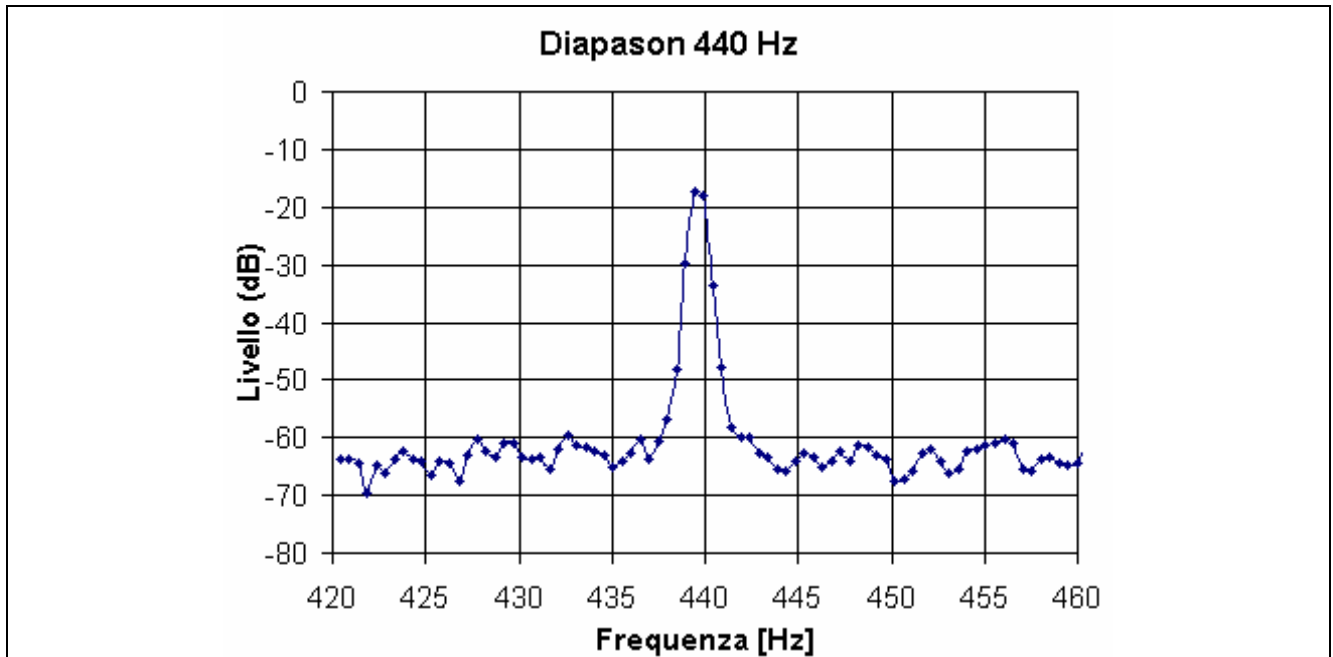

**Figura 10.** La frequenza di picco del diapason misurata (fit quadratico dei 6 punti centrali e semidifferenza delle frequenze a 50 dB) è 439.7±1.2 Hz, dall'analisi di un segnale di durata 10 s. Dallo spettro emerge anche la frequenza doppia della seconda armonica del diapason. A precisioni analoghe si è giunti con l'altro diapason campione a 442 Hz. Registratore MuVo *TX* USB2.0 Creative a frequenza di campionamento fissa a 8 kHz.

Poi ho scelto una canna di fibra di carbonio, lunga 1171.5 mm, a sezione tronco-conica con il diametro maggiore di 19 mm ed il minore di 13.3 mm, nella quale ho soffiato leggermente. Il registratore è posto lontano da altri disturbi (anche la corrente di rete disturba con i suoi 50 Hz e multipli superiori), e non sotto il flusso d'aria uscente dalla canna, per limitare il rumore. Ho soffiato dall'apertura più stretta.

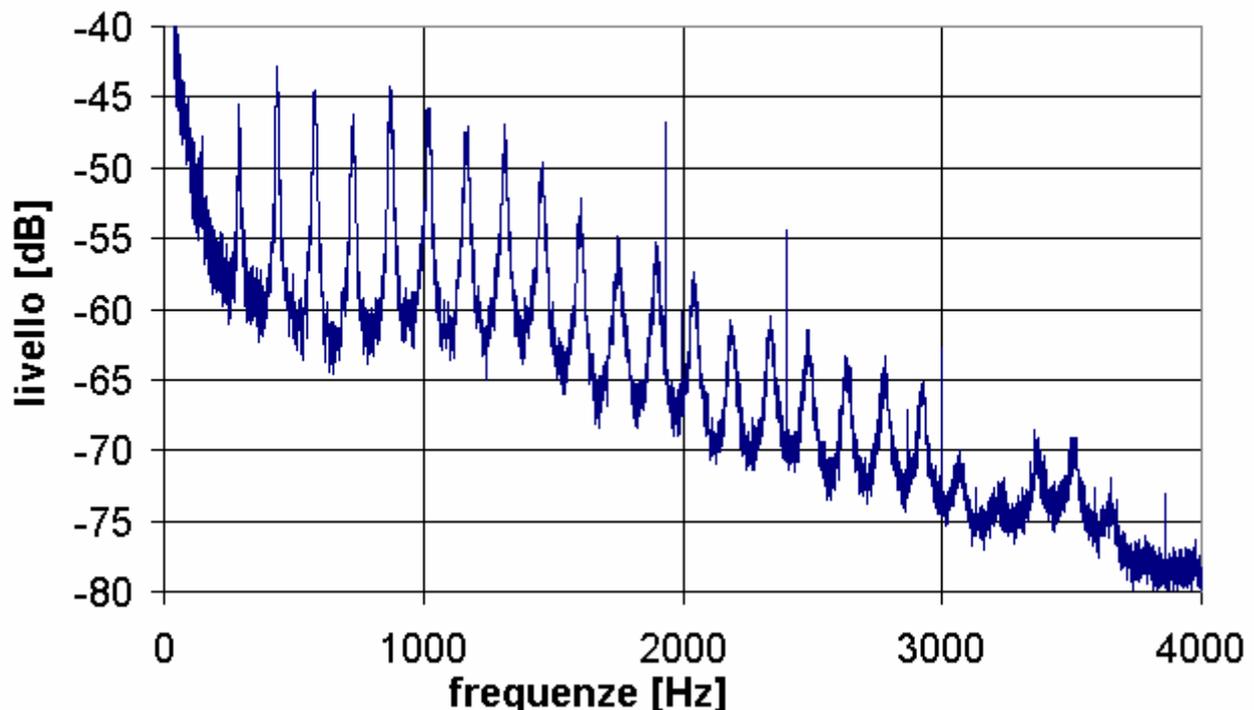

**Figura 11.** Armoniche del tubo sonoro. Ne sono state individuate 26 fino a 3789 Hz.

La correzione teorica massima α=Δ/a=0.6 deve essere applicata ai due estremi aperti della canna, e i due addendi sommati, ottenendo Δ=0.6·(8.075)mm= 4.845 mm.

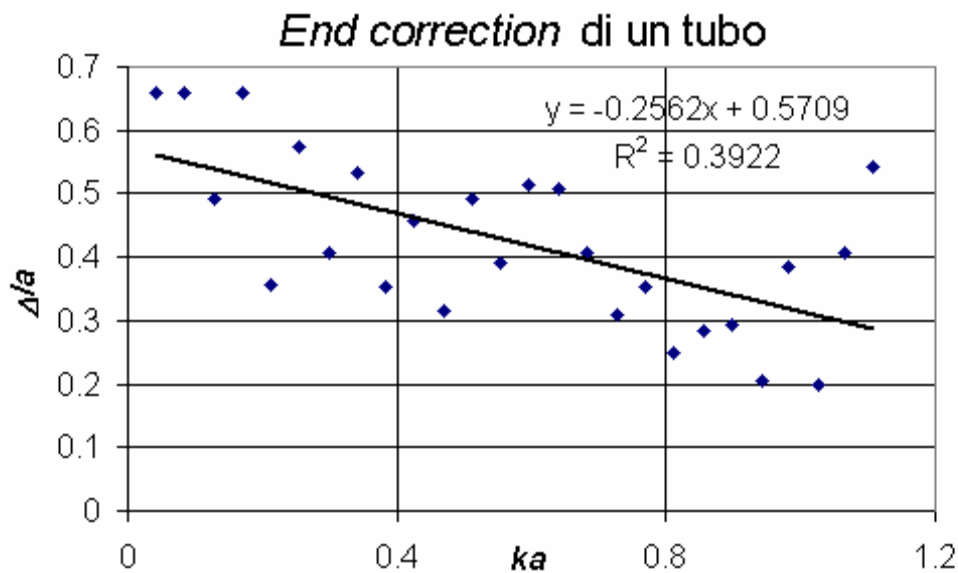

**Figura 12.** La lunghezza efficace della canna, calcolata come λ/2=c·ν viene decurtata di 1171.5 mm ed il risultato diviso per la somma dei due raggi, quindi il diametro medio (16.15 mm) ottenendo il rapporto Δ/a=α. In ascissa viene riportato il numero ka, per avere un confronto con la figura 9.
Le frequenze usate per calcolare la λ/2 sono quelle dei picchi di figura 11. La velocità del suono usata nel calcolo è c=344 m/s, essendo la temperatura T=21°C e la pressione atmosferica 1015 hPa al momento della registrazione.

Il valore teorico massimo di Δ/a=0.6 è ben riprodotto da questi dati. La retta interpolante ha una pendenza maggiore di quella teorica (2.5 volte).
Con strumenti più complessi come gli ottoni l'andamento di Δ/a si può discostare[37] di molto da quello di figura 9.
Per migliorare l'esperimento si può fare un campionamento a 48 kHz con microfoni più sensibili e spingere fino a 24 kHz (ka=6) l'analisi spettrale, mentre qui si è fermata a 3.8 kHz, il la# 7, corrispondente a ka=1.1. Tuttavia è necessario eliminare tutte le sorgenti di rumore evitando turbolenze nel flusso d'aria (soffiare piano) ed altre sorgenti sonore o disturbi della rete elettrica.

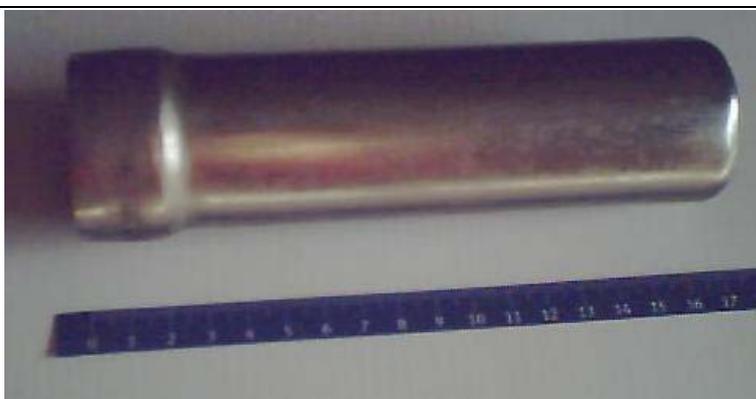

**Figura 13.** Tubo da 169 mm di lunghezza con 39 mm di diametro superiore e 46 mm di diametro inferiore. La correzione di *end* viene contata 2 volte per le due aperture e vale circa 1.2 cm per ogni apertura, mentre nel caso precedente era inferiore ai 5 mm perché il tubo era più sottile.

---

[37] Si veda l'articolo di Michael LoPresto sulle misure di *end corrections* degli ottoni usando pure gli armonici superiori.

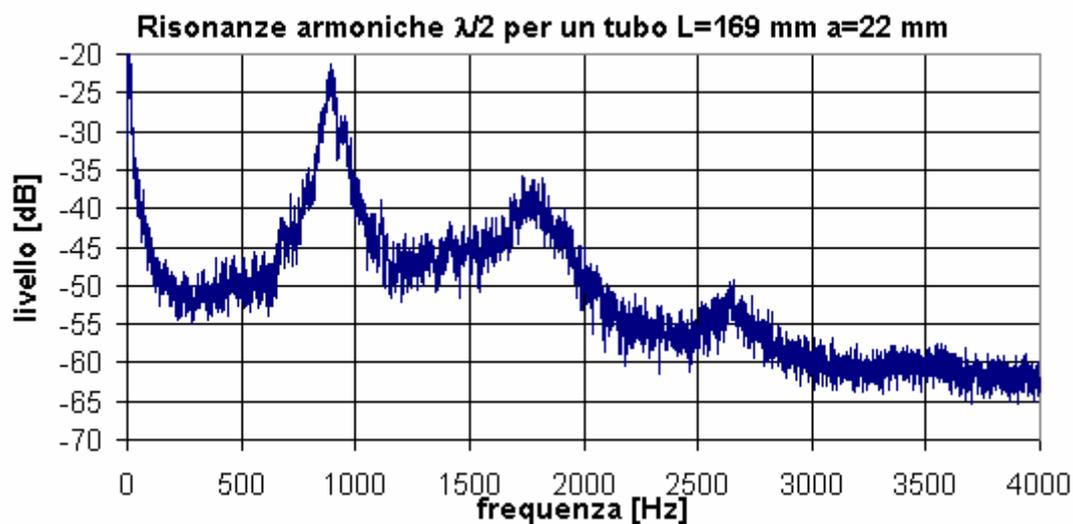

**Figura 13.** In questo tubo le risonanze sono molto più larghe del tubo precedente e di quelle tipiche degli strumenti musicali. Il suono è più sordo, anche se la frequenza della prima armonica è ben riconoscibile come un la5. Il tubo cilindrico ha due aperture di rispettivamente di 39 e 46 mm di diametro.

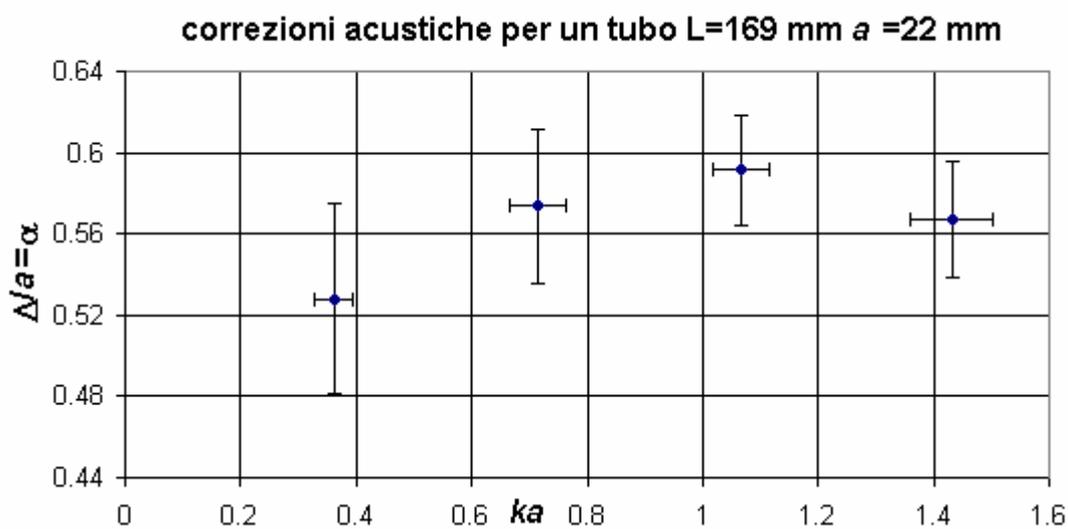

**Figura 14.** Le correzioni tengono conto anche delle ampiezze delle risonanze corrispondenti, che danno luogo alle barre d'errore. L'accordo con i valori teorici previsti è ancora buono.

Le correzioni acustiche di *end* tendono a ridursi con l'aumentare della frequenza, tuttavia il valore di α lo si può considerare costante nell'ambito delle ottave del pianoforte. Quindi anche nel caso delle ottave di Gerberto la correzione in α sarebbe stata la stessa per le due ottave. Gerberto però considerava i rapporti tra le differenze di lunghezze e, soprattutto, ha ricavato la lunghezza della canna più lunga dell'ottava inferiore, con una proporzione geometrica. Quindi ha cercato la legge unificatrice che spiegasse le stesse correzioni, piuttosto che continuare l'applicazione delle correzioni empiriche anche all'ottava inferiore.

## *Correzioni di bocca*

Quanto alle correzione di bocca, occorre necessariamente procedere con la *fistula* aperta dai due lati, e confrontare i risultati con quella chiusa alla fine, senza *end correction*.
Il flauto dolce soprano, quello che si usa ancora nell'educazione musicale alle scuole medie, è lo strumento che, con tutti i fori tonali chiusi, meglio riproduce la struttura base di una canna d'organo.

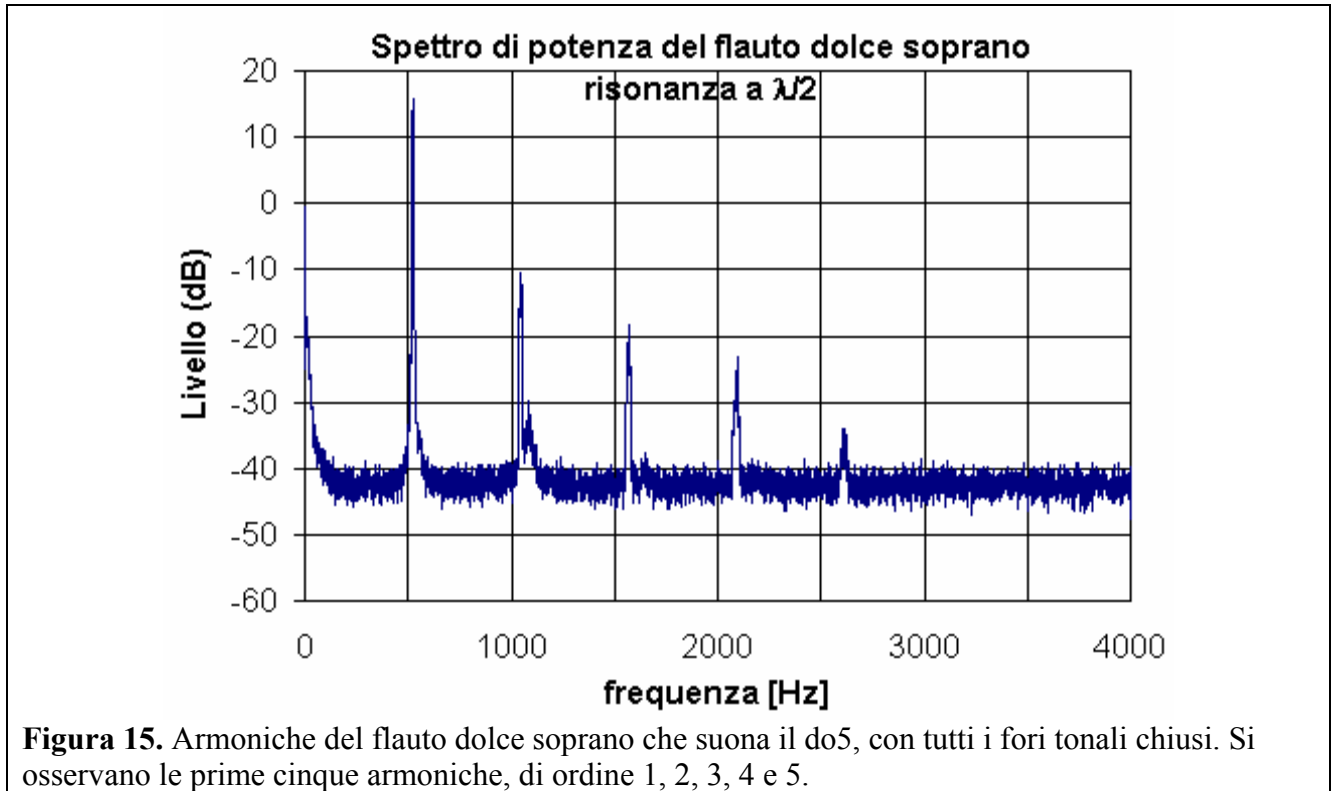

**Figura 15.** Armoniche del flauto dolce soprano che suona il do5, con tutti i fori tonali chiusi. Si osservano le prime cinque armoniche, di ordine 1, 2, 3, 4 e 5.

Se si chiude anche l'apertura finale del tubo si ottiene un mi4, ed il flauto risuona a λ/4.

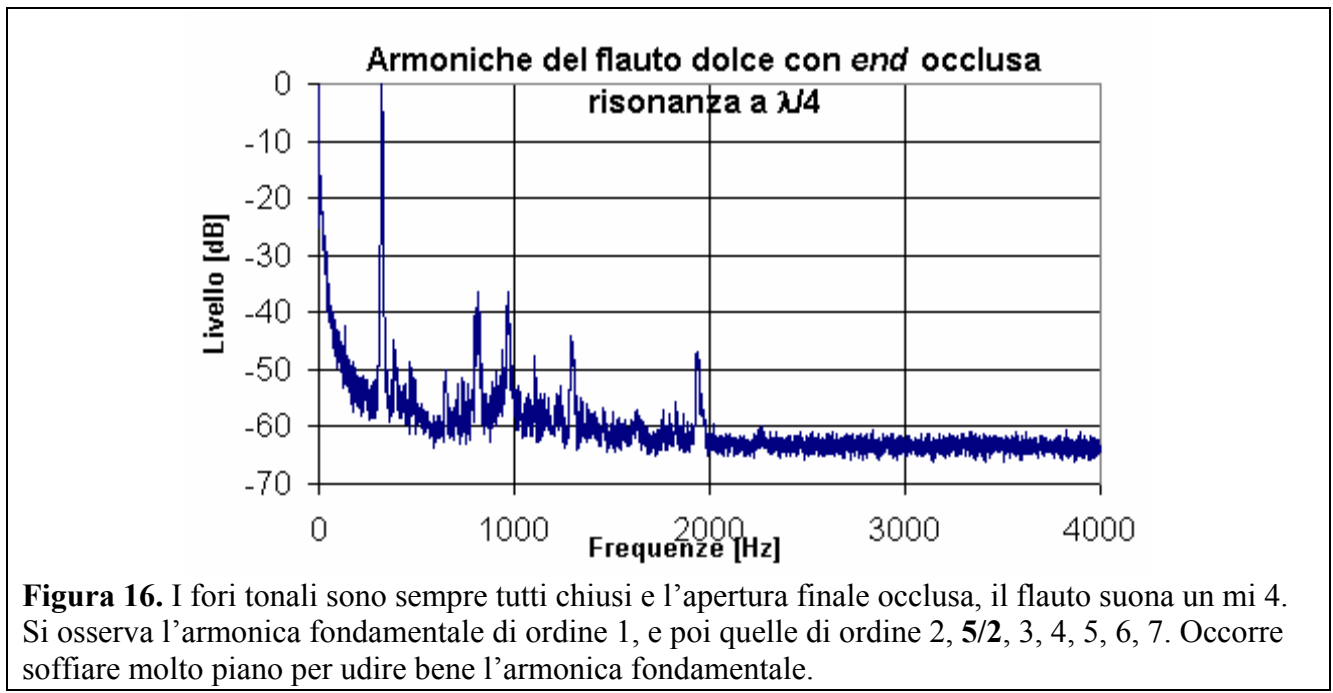

**Figura 16.** I fori tonali sono sempre tutti chiusi e l'apertura finale occlusa, il flauto suona un mi 4. Si osserva l'armonica fondamentale di ordine 1, e poi quelle di ordine 2, **5/2**, 3, 4, 5, 6, 7. Occorre soffiare molto piano per udire bene l'armonica fondamentale.

Dalle frequenze misurate si ricavano i valori di λ/2 e λ/4 con cui risuona il flauto, e si sottraggono i secondi ai primi, ottenendo la correzione di bocca.

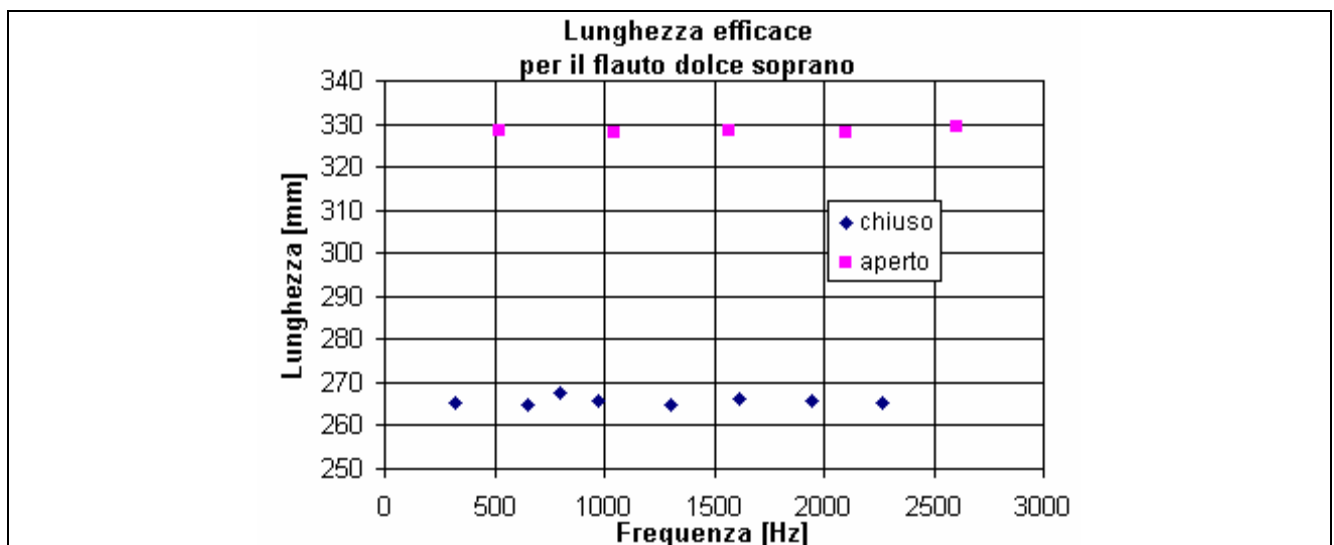

**Figura 17.** Utilizzando tutte le armoniche misurabili con il flauto dolce, e considerando l'ordine delle armoniche, si è ricavata per ciascuna armonica la lunghezza efficace del flauto. La lunghezza efficace del flauto aperto normalmente risulta 328.4±0.6 mm, mentre di quello con l'apertura occlusa vale 265.8±0.9. La correzione di bocca vale 62.6±1.1 mm.

La misura di 62.6 mm corrisponde esattamente con la distanza tra il becco e la fine della scanalatura superiore, come si vede nella foto seguente.

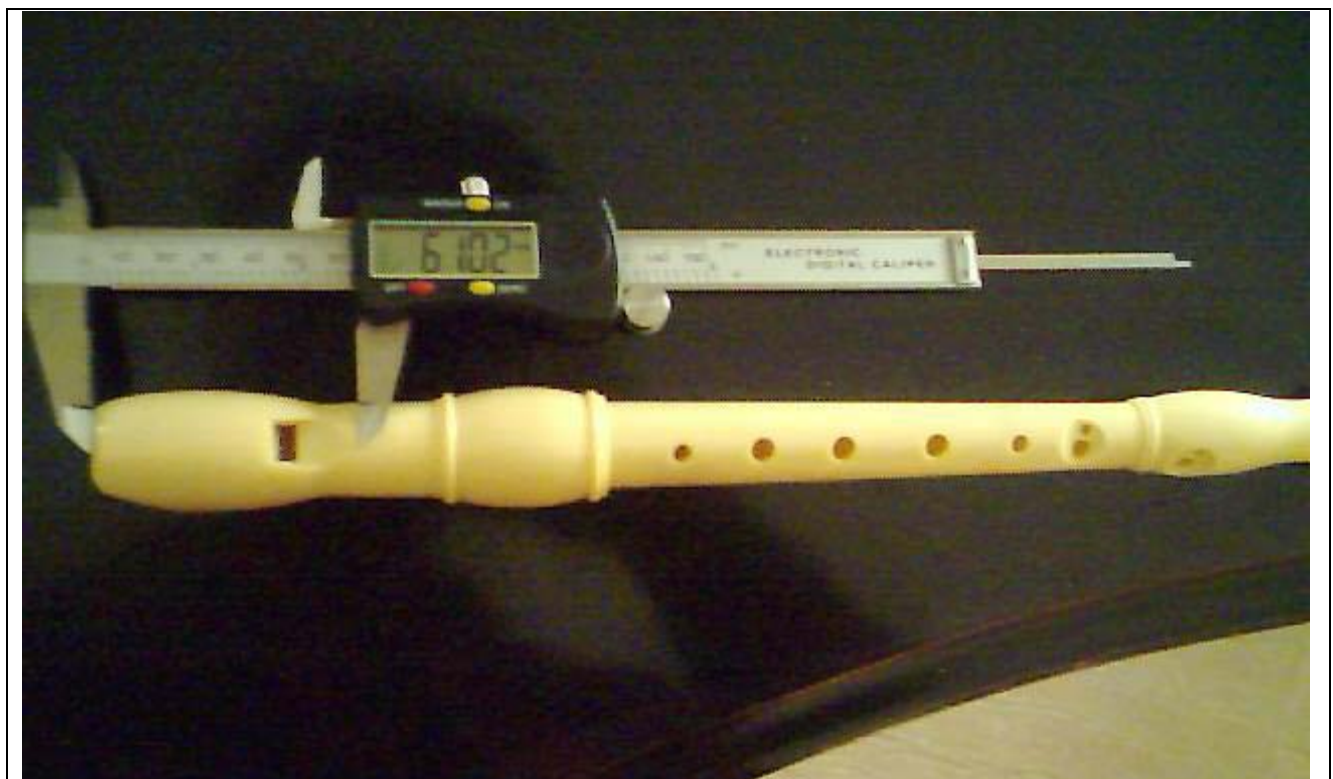

**Figura 18.** La distanza calcolata è 62.6±1.1 mm e quella misurata è 61.02 mm. In ottimo accordo.

Dunque nelle canne d'organo è ragionevole aspettarsi una correzione di bocca pari alla *statio e la flexura* assieme, ed è verosimile che le correzioni usate da Gerberto non riguardino quindi il solo *corpus fistulae*.

## *Conclusioni*

Gerberto, compositore[38] e musico egli stesso,[39] era ben a conoscenza che le canne d'organo erano di diametri differenti. Infatti spiegando a Costantino di Fleury l'uso di *fistulae* per puntare la stella Polare da montare all'interno della sfera afferma che le *fistulae* per queste osservazioni differiscono da quelle dell'organo, essendo tutte uguali in dimensione, al fine di non distorcere la visione di alcuno che osservi i circoli nei cieli.[40]

Richero di Reims (Historia Francorum III, 49) riporta che Gerberto usava il monocordo per insegnare le differenze tra i generi musicali.[41]

Inde etiam musicam, multo ante Galliis ignotam, notissimam effecit. Cujus genera in monocordo disponens, eorum consonantias sive simphonias in tonis ac semitoniis, ditonis quoque ac diesibus distinguens, tonosque in sonis rationabiliter distribuens, in plenissimam notitiam redegit.

*Poi insegnò anche la musica, fino ad allora del tutto ignorata in Gallia. Rese infatti tale disciplina comprensibile e razionale, disponendo secondo il monocordo i vari modi, classificando le loro consonanze in toni e semitoni, distinguendole anche in ditoni e diesis, e suddividendo matematicamente i toni in suoni.[42]*

Dunque il suo trattato sulla *Mensura Fistularum* si configura anch'esso come un trattato didattico, in cui si vuole mostrare la commensurabilità, ovvero la subordinazione a medesime leggi fisiche, tra le *fistulae* ed il monocordo.

In questo senso l'opera di Gerberto costituisce per la *Musica Instrumentalis* un trait-d'union con la *Musica Mundana* regolata dai numeri pitagorici. È l'equivalente dell'equante tolemaico che *salva i fenomeni* -sempre meglio osservabili-, complicando il modello iniziale basato su epicicli e deferenti, ma tenendo fermo il principio dei moti circolari uniformi.
I moltiplicatori 13 ½ e 14+1/3+1/144+1/288, sostituiscono il numero 12 valido per il monocordo e riconducono le canne d'organo nello stesso alveo ermeneutico, quello della teoria musicale pitagorica.
La soluzione trovata da Gerberto è ingegnosa e laboriosa, ma ottiene precisamente il suo scopo: *fistulae* e monocordo possono essere trattati con gli stessi strumenti matematici, a patto di cambiare dei parametri. Una visione piuttosto moderna di una branca della fisica, quella delle correzioni acustiche alle lunghezze dei tubi sonori, che usa ancora oggi molti dati empirici a completamento dei modelli teorici troppo semplificati e pur già estremamente complessi.
Questo trattato non era concepito per la costruzione di un organo, come hanno già osservato vari commentatori, ma servì certamente a favorirne la diffusione in ambito liturgico *ex auctoritate Domino papae Gerberto*.[43]
Gerberto *Rogatus a Pluribus* ha messo per iscritto, in uno dei suoi pochi trattati, un argomento che costituisce uno spartiacque tra il pensiero scientifico antico e quello moderno: un problema di *Musica Instrumentalis*, viene affrontato con gli stessi metodi delle più nobili forme di *Musica*, quello della ricerca di regolarità matematiche, e mantenendo allo stesso tempo un legame con i dati empirici.
Dietro c'è la ricerca, da parte di Gerberto, dell'unità razionale del Creato,[44] dell'impronta stessa del Creatore che si manifesta in un linguaggio matematico: una prospettiva certamente pitagorica e neoplatonica, ma tutt'affatto lontana da quella che sarà di Galileo.

---

[38] Gerberto ha composto un ufficio per Saint-Géraud, Paris Bibl. Nazionale ms. lat. 944 e 2826. P. Riché, *Il Papa dell'Anno Mille –Silvestro II*, Cinisello Balsamo, 1988 p. 55, nota 16. Olleris (1867) riporta anche di un inno allo Spirito Santo. Michel Huglo (2001 nota 23 a p. 227-228), il musicologo che attribuì a Gerberto l'ufficio per Saint-Géraud ha poi ritrattato questa attribuzione.

[39] Nella lettera 105 a Bernardo nella numerazione di H. Pratt Lattin (1961), 92 in quella di J. Havet, Michel Huglo vuole leggere ivi una dichiarazione che egli stesso sapeva suonare l'organo, ma non bene M. Huglo. *Gerberto Teorico Musicale, visto dall'anno 2000*, Bobbio 2001. Tuttavia il testo latino non consente questa interpretazione e lì Gerberto dice che non è in grado di completare un compito che potrà fare Costantino di Fleury (così traduce anche la Pratt Lattin).
F. G. Nuvolone *Gerberto e la* Music,a ABobSt **5** (2005) che Gerberto si sia applicato non solo all'insegnamento quale *musicus*, ma pure all'uso liturgico dell'organo e al *cantus planus*.

[40] Lettera a Costantino di Fleury, n. 2 nella numerazione di H. Pratt Lattin (1961).

[41] È probabile che Gerberto abbia dimostrato che non si doveva paragonare la divisione del monocordo a quella delle canne d'organo in un trattato sul monocordo, come scrive P. Riché, *Il Papa dell'Anno Mille –Silvestro II*, Cinisello Balsamo, 1988 p. 55; e solo successivamente *Rogatus a Pluribus* abbia trovato la chiave di volta che metteva i due strumenti d'accordo?

[42] Traduzione di L. C. Paladino, op. cit. (2006).

## *Ringraziamenti*

---

[43] Così Bernellino nella prefazione al De Abaco indirizzato ad un certo Amelio dichiara di dovere tutto al Signor Papa Gerberto. Da P. Riché, *Il Papa dell'Anno Mille –Silvestro II*, Cinisello Balsamo, 1988 p. 236. A questa conclusione giunge anche Erminia Santi, *Gerberto e la Musica*, Roma (2003).
[44] F. G. Nuvolone (ABob **26** 2004) per descrivere il pensiero di Gerberto riporta la citazione di Isidoro di Siviglia (*Etimologie* III, 4,3) che la natura per l'atto creativo di Dio è passata dal chaos inarticolato ad essere interamente disposta secondo numero, misura e peso.

## *Bibliografia*